\documentclass[twocolumn,amsmath,amssymb,superscriptaddress,footinbib]{revtex4-2}

\usepackage{graphicx}
\usepackage{hyperref}
\usepackage{braket}
\usepackage{bbm}
\usepackage{layouts}
\usepackage{siunitx}

\usepackage[english]{babel}
\usepackage[svgnames]{xcolor}
\usepackage{ulem}
\begin{document}
\rmfamily
\raggedbottom

\title{Programmable control of the spatiotemporal quantum noise of light}

\newcommand{\MITphys}{Department of Physics, Massachusetts Institute of Technology, Cambridge, MA 02139, USA.}
\newcommand{\MITRLE}{Research Laboratory of Electronics, Massachusetts Institute of Technology, Cambridge, MA 02139, USA.}
\newcommand{\Stanford}{E. L. Ginzton Laboratory, Stanford University, Stanford, CA 94305, USA.}
\newcommand{\Technion}{Department of Electrical and Computer Engineering, Technion – Israel Institute of Technology, 32000 Haifa, Israel.}
\newcommand{\Harvard}{Department of Physics, Harvard University, Cambridge, MA 02138, USA.}
\newcommand{\Cornell}{School of Applied and Engineering Physics, Cornell University, Ithaca, New York 14853, USA. \\ $^\dagger$ Denotes equal contribution.}

\author{Jamison Sloan$^\dagger$}
\affiliation{\MITRLE}
\affiliation{\Stanford}

\author{Michael Horodynski$^\dagger$}
\affiliation{\MITphys}

\author{Shiekh Zia Uddin$^\dagger$}
\affiliation{\MITphys}

\author{Yannick Salamin}
\affiliation{\MITphys}

\author{Michael Birk}
\affiliation{\Technion}

\author{Pavel Sidorenko}
\affiliation{\Technion}

\author{Ido Kaminer}
\affiliation{\Technion}

\author{Marin Solja\v{c}i\'{c}}
\affiliation{\MITRLE}
\affiliation{\MITphys}

\author{Nicholas Rivera}
\affiliation{\Harvard}
\affiliation{\Cornell}
\email{E-mail: jsloan@stanford.edu, mhorodyn@mit.edu, nrivera@cornell.edu}

\begin{abstract}
Optoelectronic systems based on multiple modes of light can often exceed the performance of their single-mode counterparts. However, multimode nonlinear interactions often introduce considerable amounts of noise, limiting the ultimate performance of these systems. It is therefore crucial to develop ways to simultaneously control complex nonlinear interactions while also gaining control over their noise. Here, we show that noise buildup in nonlinear multimode systems can be strongly suppressed by controlling the input wavefront. We demonstrate this approach in a multimode fiber by using an active wavefront-shaping protocol to focus a region of high intensity --- yet low intensity noise --- at the output. Our programmable control of both the input and output reduces the beam noise by 12 dB beyond what linear attenuation achieves, reaching levels near the quantum shot-noise limit. We show that this is possible because the optimally shaped wavefront maximally decouples the output intensity fluctuations from the input laser fluctuations. These findings are supported by a new theoretical and simulation framework that efficiently captures spatiotemporal quantum noise dynamics in highly multimode nonlinear systems. Our results highlight the potential of programmable wavefront shaping to enable nonlinear multimode technologies that overcome noise buildup to operate at quantum-noise limits.
\end{abstract}

\maketitle

\begin{figure*}[ht!]
    \includegraphics{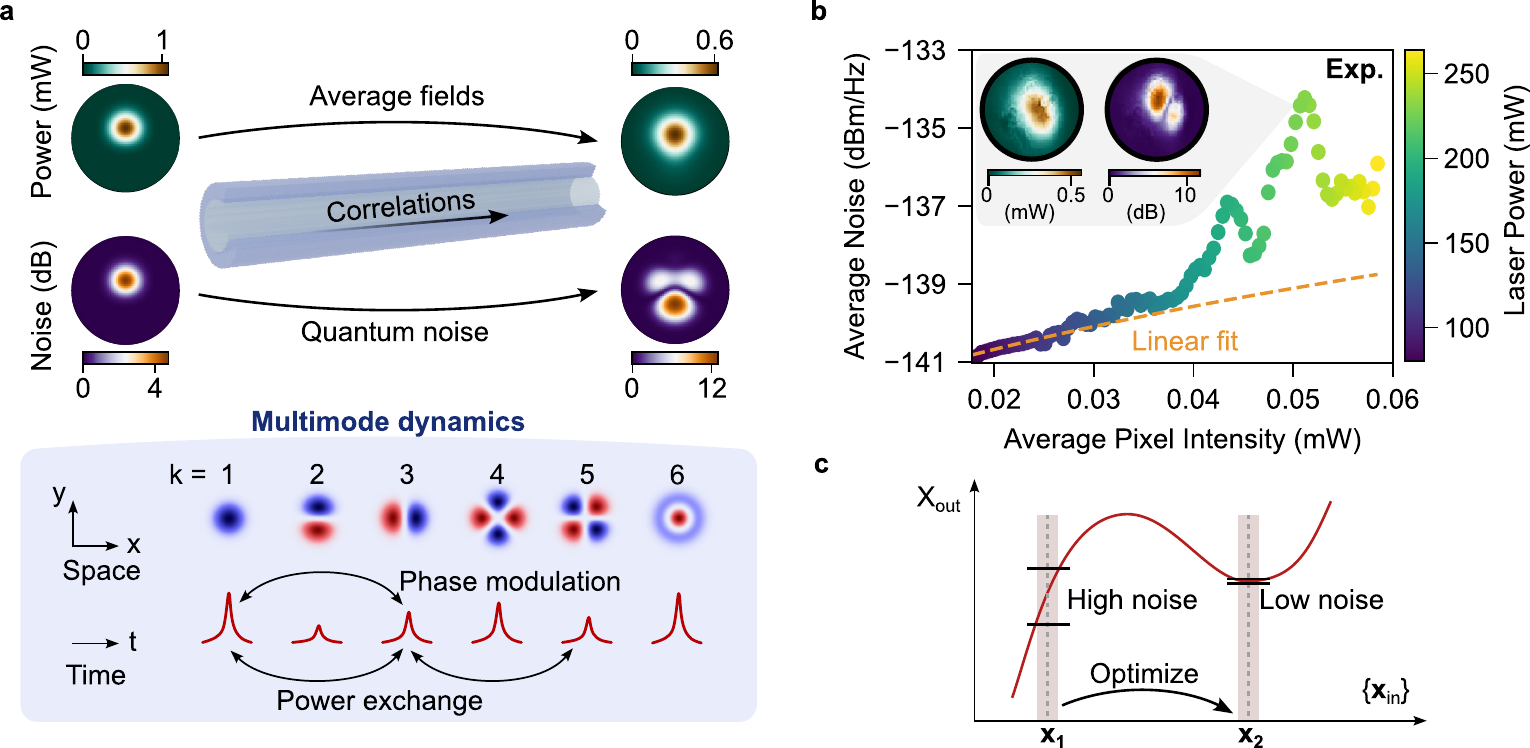}
    \caption{\textbf{Noise propagation in nonlinear multimode photonics.} (\textbf{a}) (Top) Schematic of ultrafast pulses propagating in a multimode fiber. Simulated images show how the intensity and intensity noise profile of the beam at the input and output of the fiber are shaped through multimode nonlinear interactions. The noise is characterized relative to a set noise floor level, which allows for the closest comparison to experiments. (Bottom) Representation of the degrees of freedom in a multimode fiber: spatial modes (shown with modal profiles) and temporal pulse profiles interact through Kerr nonlinearity, as well as other effects such as Raman scattering. (\textbf{b}) Experimental results showing how noise proliferates with increasing nonlinear interactions between many degrees of freedom. For different pixels $i$ of the output beam, we measure the intensity and intensity noise (specified in dBm/Hz). Averaging over the pixel intensity and noise at different input power levels shows how the noise level generally increases with increasing input power. The dashed orange line shows the expected noise growth in a linear system. The inset shows an example of the intensity and noise images obtained at one particular power. (\textbf{c}) Noise propagation in nonlinear multimode photonics depends heavily on the initial condition (e.g., initial modal pulse profiles), which are collected into a vector of initial conditions $\mathbf{x}_{\text{in}}$. For many initial conditions (e.g. $\mathbf{x}_1$), the noise in some measured quantity $X_{\text{out}}$ may be quite large, due to the uncontrolled interactions between many degrees of freedom. However, other initial conditions (e.g. $\mathbf{x}_2$) may spread much less noise to $X_{\text{out}}$.}
    \label{fig:schematic}
\end{figure*}

The invention of the laser has led to the generation of highly coherent and intense light. These laser sources have provided access to photonic nonlinearities \cite{hecht_how_2010}, which have enabled a wide range of applications, including frequency conversion \cite{franken_generation_1961}, ultrashort pulse formation \cite{keller_recent_2003}, and enhanced microscopy \cite{denk_two-photon_1990,hell_breaking_1994,freudiger_label-free_2008}. Often, these functionalities can be improved by employing multiple modes of light (e.g., frequency, spatial mode, polarization) in platforms such as free-space optics, fibers, and integrated nanophotonics \cite{wright_physics_2022,lu_frequency-bin_2023,puttnam_space-division_2021,wang_orbital_2022,wright_nonlinear_2022,vitali_nonlinear_2025}. The high dimensionality of these so-called ``multimode'' systems makes them attractive for applications including carrying large amounts of power \cite{chen_suppressing_2023,rothe_wavefront_2025}, as well as information \cite{richardson_space-division_2013}. These systems also exhibit complex phenomena including soliton formation \cite{haus_mode-locking_2000}, optical thermalization \cite{wu_thermodynamic_2019,pourbeyram_direct_2022}, spatiotemporal pulse dynamics \cite{wright_spatiotemporal_2017}, and mimicking neural networks \cite{wright_deep_2022,momeni_backpropagation-free_2023}. Additionally, recent works have begun to explore the quantum optical behavior of these systems \cite{guidry_quantum_2022, presutti_highly_2024, lustig_quadrature-dependent_2025,zia_uddin_noise-immune_2025}. In this context, the scaling of complexity with the number of modes is yet larger, making multimode nonlinear platforms strong candidates for quantum information \cite{cai_multimode_2017,asavanant_generation_2019,barral_versatile_2020} and metrology \cite{nichols_multiparameter_2018}.

A critical aspect for many applications of light is noise, characterized by random deviations of an optical field from its average value. Noise limits many important use cases, including interferometry \cite{the_ligo_scientific_collaboration_gravitational_2011}, microscopy \cite{brida_experimental_2010}, spectroscopy \cite{whittaker_absorption_2017}, computing \cite{ma_quantum-limited_2025} and information processing \cite{shapiro_quantum_2009,weedbrook_gaussian_2012}. Due to quantum effects, even perfectly coherent light from an ideal laser carries ``shot noise'' associated with the ``standard quantum limit'' (SQL) \cite{loudon_quantum_2000}. Shot noise leads to Poissonian fluctuations in the intensity of light, as well as fluctuations in relative phase that go inversely with the number of photons in the light. Moreover, many light sources --- especially intense ones --- operate at noise levels well above the SQL. For example,  many high-power laser sources exhibit excess noise due to spontaneous emission, as well as other noise sources \cite{bachor_guide_2019}. As another example, intense white light ``supercontinuum" sources suffer from large noise due to the amplification of shot noise by the nonlinear dynamics, which widens the frequency spectrum of incoming pulses \cite{corwin_fundamental_2003}. In general, one expects that uncontrolled nonlinear interactions between many modes will cause noise to accumulate in any individual mode.

Despite its importance, many questions remain open regarding noise behavior in the highly multimodal regime. In particular, it is not well understood how spatiotemporal nonlinearities transform input noise into fluctuations of different output quantities. As a result, strategies to mitigate multimode noise buildup and ideally suppress output noise to the quantum level have yet to be developed. This work addresses the central question of whether nonlinear multimode interactions can be controlled to overcome noise amplification and instead produce low-noise light.

We show that controlling the wavefront of light incident on a nonlinear multimode system can suppress fluctuations in target quantities down to the quantum-noise limit, even when starting with highly noisy sources. While on average, unstructured multimodal interactions amplify fluctuations, we show that specific input phase profiles reduce noise by more than 12 decibels compared to linear attenuation. We further demonstrate that noise and intensity can be co-optimized (without a detailed understanding of the nonlinear dynamics): in some regions, the intensity remains fixed while the noise falls by nearly a factor of 20, to levels near the shot noise limit. Using a theory and simulation framework we developed for spatiotemporal quantum noise dynamics, we find that cross-phase modulation generally drives noise build-up. With appropriate wavefront shaping, however, the system can be steered so that the output is largely insensitive to excess input noise. In this case, the output has fluctuations near the shot-noise level, even though the input light is far noisier. We demonstrate these effects in the context of femtosecond pulses propagating in spatially multimode fibers, but these effects are general to a wide variety of nonlinear spatiotemporally multimode systems. 

\begin{figure*}[ht!]
    \centering
    \includegraphics{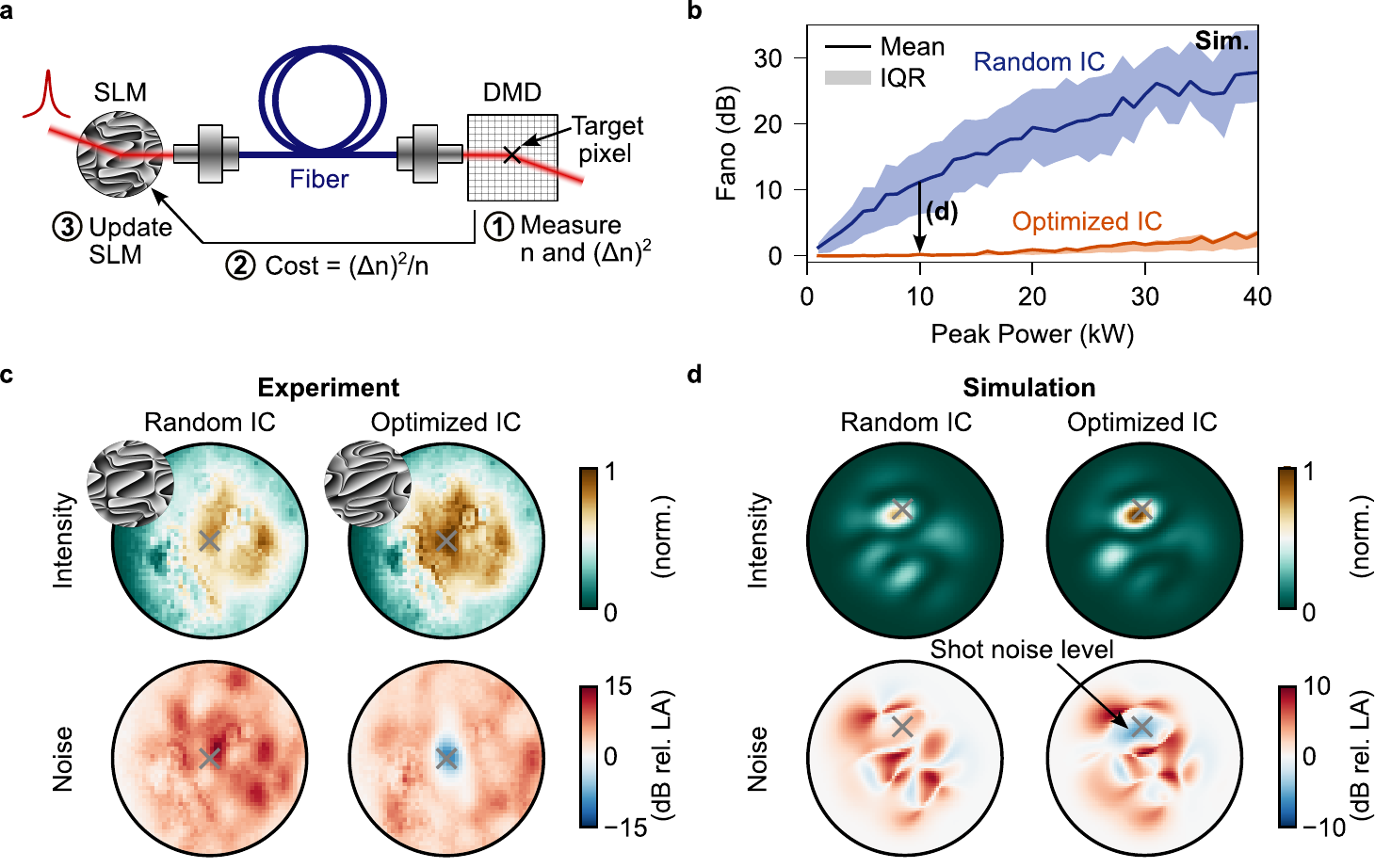}
    \caption{\textbf{Optimization of noise with wavefront shaping.} (\textbf{a}) To lower the intensity noise of the beam at the center of the output facet of the fiber, we iteratively change the modal composition of the incoming light with a spatial light modulator by using the measured intensity and noise as feedback. (\textbf{b}) Simulated optimization of noise in a small region (pixel) of the beam by tuning the initial condition, for different input power levels. Noise is characterized by the Fano factor $F = (\Delta n)^2/n$, which normalizes the variance in photon number to the shot noise level. IQR denotes the interquartile range in which half of the data falls. (Lower) Comparison of the spatial intensity and noise maps before and after the optimization procedure for experimental (\textbf{c}) and simulated (\textbf{d}) data, respectively. The corresponding SLM patterns in the experiment are shown as an inset. The noise is shown in dB relative to linear attenuation of the whole beam to the observed power level (rel. LA). We mark the target pixel for optimization with a cross.}
    \label{fig:2}
\end{figure*}

\subsection*{Noise propagation in multimode nonlinear photonics}

We now introduce the multimode optical fiber, which we take as an example system for studying the propagation of noise in the presence of multimode nonlinear dynamics (Fig.~\ref{fig:schematic}a). Here, the optical field is described by ultrafast optical pulses propagating in different spatial modes. As is well known, the intensity-dependent refractive index (i.e., Kerr effect) leads to nonlinear mixing of both frequency and spatial degrees of freedom by mechanisms such as self- and cross-phase modulation, as well as modal power exchange. The simulation framework we developed allows us to compute both the average field and noise properties of ultrafast pulses propagating through multimode fibers. An example result for a 210 fs pulse incident on the fiber with a slight spatial offset is shown in Fig.~\ref{fig:schematic}a. While the pulse shape is slightly modified, the intensity noise is completely reshaped by the nonlinear interaction, exhibiting a spatial pattern that no longer matches the beam shape. Moreover, the nonlinear interactions amplify the noise peak by nearly an order of magnitude relative to the input state. 

Based on this concept, we developed an experimental platform that allows us to observe multimode noise dynamics. At the output of the fiber, we characterize the spatial intensity distribution (i.e., image of the beam), as well as the distribution of intensity noise across the beam (i.e., the variance in the intensity at different spatial points). Here, this variance is captured by the power spectral density of the current generated by a photodiode at some frequency for which technical noise sources are minimal (10 MHz in our experiment) \footnote{We consider the power spectral density at a given frequency rather than the frequency-integrated power spectral density (which is proportional to the photon number variance in a pulse), as this allows us to make direct comparisons with the quantum shot noise level (as is standard). In contrast, the integrated power spectral density will typically be far from the shot noise level due to low-frequency technical noise sources --- however this noise can also be reduced by the same methods described here due to the quasi-instantaneous nature of Kerr nonlinearity.}. For light with intensity fluctuations at the shot-noise level, this power spectral density is directly proportional to the intensity. Such characterization confirms that nonlinear fiber dynamics can cause the intensity noise of the beam to take on a different spatial profile than the intensity itself (Fig.~\ref{fig:schematic}b, inset). The impact of the nonlinearity on the noise can further be seen by computing the average intensity and noise across the whole beam for different input power levels (Fig.~\ref{fig:schematic}b). With increasing input power, the average intensity noise level across the pixels grows compared to what is expected from a linear system. 

This type of noise amplification is what is generally expected from a nonlinear system with many degrees of freedom. A generic nonlinear dynamical system consists of some initial conditions (ICs) $\mathbf{x}_{\text{in}}$. In our system, these are the complex values of the spatiotemporal electric field incident on the input facet of a fiber. Some quantity of interest $X_{\text{out}}$ (in our experiments, the intensity of a pixel) is then measured at the output. This quantity will carry some noise, characterized by $(\Delta X_{\text{out}})^2$, which originates from fluctuations in the initial conditions (from technical noise, or even quantum fluctuations). For many initial conditions (e.g., $\mathbf{x}_1$), input fluctuations are amplified --- consistent with the sensitivity of many nonlinear optical phenomena to initial conditions \cite{cavalcanti_noise_1995,caves_quantum_1982,walmsley_observation_1983,dudley_supercontinuum_2006,gross_superradiance_1982,roques-carmes_biasing_2023,pellegrini_physics_2016}. The key concept behind this work is that control over the input fields can allow one to find optimized initial conditions ($\mathbf{x}_2$) which generate much less noise at the output. In this case, the average noise-amplifying properties of multimode nonlinearities are overcome; instead, the nonlinear dynamics are harnessed as a resource that can suppress even large amounts of input noise. 

\subsection*{Optimization of noise via input shaping}

We now demonstrate that optimization of initial conditions can steer a nonlinear system into an output state exhibiting low noise for a chosen quantity of interest. In our experiments, our quantity of interest is the intensity at the center of the fiber. We implement this by directing the beam from the output facet of the fiber to a digital micromirror device (DMD), enabling the photodiode to detect light from a single pixel (or an arbitrary combination thereof). The initial conditions are controlled by wavefront shaping \cite{cao_shaping_2022}, achieved via a spatial light modulator (SLM) on the input side of the fiber. This reconfigurable phase mask allows us to modify the populations and phases of spatial modes excited at the input of the fiber. Then, combining the SLM input with an optimization algorithm allows us to steer the system toward a desired noise behavior (Fig.~\ref{fig:2}a). Similar approaches have been used to control spectral and spatial properties of the mean-field in nonlinear (multimode) fibers \cite{tzang_adaptive_2018,florentin_shaping_2016}. Wavefront shaping has also been recently employed to mitigate nonlinear instabilities \cite{chen_suppressing_2023,wisal_optimal_2024,boussafa_deep_2025} and stimulated Brillouin scattering \cite{chen_mitigating_2023,wisal_theory_2023,rothe_wavefront_2025} in multimode fibers for high power applications. The general operating principle of the preceding studies is to distribute the power efficiently between many modes to suppress the rate of noise amplification. In contrast, our work aims to mitigate excess noise and reduce the noise of a light beam down to the quantum level.

In the experiment, we begin with a random IC (obtained through a random SLM pattern). Initially, both intensity and noise show a speckle pattern, a hallmark sign of random scattering within the MMF (see Fig.~\ref{fig:2}c). This random scattering arises from deliberate stressing of the fiber. We found that enhanced random mode coupling aids us in the optimization procedure, as it increases the space of nonlinear dynamics that can take place in the fiber. The intensity noise level for each pixel is shown in comparison to the noise level found by linear attenuation (LA) of the initial laser source down to the observed pixel power. We choose this method of comparison because the source laser carries a high degree of excess noise (compared to the shot noise level), which is reduced by linear attenuation. Thus, assessing the noise relative to linear attenuation allows us to isolate the effect of nonlinearity on the noise. This method of analysis reveals that for the random IC, the beam exhibits large intensity fluctuations, with nonlinear interactions contributing 10-15 dB of additional noise. 

We then choose the center pixel as the target quantity to optimize for high intensity, but low noise. To achieve this, we use the intensity-normalized noise as a cost function, and then perform gradient-free optimization on the physical system, adjusting the SLM pattern at each optimization step. After 100 optimization steps, we found ICs that dramatically lower the noise for the target pixel, while maintaining a similar intensity (see Fig.~\ref{fig:2}c). Noise imaging of the optimized state reveals a pronounced low noise region in the center of the beam, where the intensity fluctuations are more than 10 dB below the limit set by linear behavior, and a 20 dB improvement compared to the initial random state.

Our experimental demonstration is well supported by optimization performed on simulations. For these optimizations, we simulated continuous wave propagation due to the high computational overhead needed to optimize spatiotemporal simulations. Our main conclusions are, however, not impacted, since (1) the lower bound for noise minimization, which is approximately given by the shot noise, is the same for continuous and pulsed waves; and (2) the experimentally measured spectrum shows limited spectral dynamics (see Supplementary Information (SI)). We also note that our simulations do not include random mode coupling. Since we can create arbitrary initial conditions, this enables the simulated system to explore a larger space of dynamics than available in our experiment. Fig.~\ref{fig:2}b shows the simulated noise of various target pixels as a function of total input power. Noise values are reported relative to the standard quantum limit, characterized by the Fano factor $F = (\Delta n)^2/n$. Indeed, random ICs typically result in noisy pixels ($F \gg 1$) at the output due to the unmitigated growth of fluctuations during the nonlinear propagation. By performing gradient optimization on the simulations, we find that noise reduction by shaping the input conditions is a robust capability that occurs across many input powers and many target pixels. Fig.~\ref{fig:2}d shows the results of one particular optimization run as a comparison to the experiment. Here, the optimization procedure drops the noise by nearly an order of magnitude down to the shot noise level, while also slightly improving the intensity of the focused region.

Our theory provides additional insight into the ability of wavefront shaping to drastically reduce the spread impact of excess input noise. Specifically, we derive (see SI) an equation for the variance in the photon number $n$ for a particular pixel:
\begin{align}
    \left( \Delta n \right)^2 & = n \left( 1 - \Phi \right) + \sum_{m}^M \left| \frac{\partial n }{\partial u_{m}^{(0)}} \right|^2 +  \delta F_\mathrm{in} \left| \sum_{m}^M U_{m} \frac{\partial n }{\partial u_{m}^{(0)}} \right|^2.
    \label{eq:varn}
\end{align}
Here, $\Phi$ is a small mode-dependent factor ($\sim10^{-2}$), $\partial n / \partial u_m^{(0)}$ are the sensitivities with respect to the initial modal fields $u_m^{(0)}$, $\delta F_\text{in}$ quantifies the excess input noise in the laser and $U = u^{(0)}/|u^{(0)}|$ is the normalized input. For random ICs, the last term in Eq.~\eqref{eq:varn} describing excess noise dominates, while for optimized ICs it is suppressed by over an order of magnitude. This leads us to conclude that our optimization procedure works by removing the influence of excess input noise on the output noise.

\begin{figure*}[t!]
    \centering
    \includegraphics{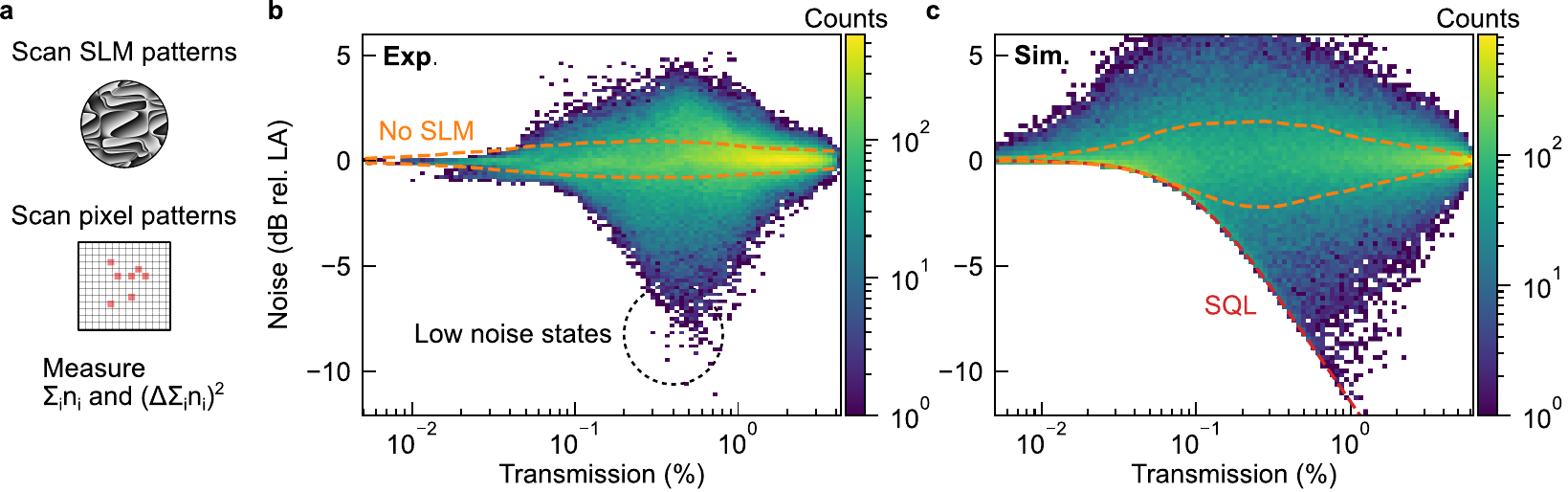}
    \caption{\textbf{Optimal control of the input phase profile and the output spatial filter.} 
    (\textbf{a}) Noise of the output depends both on the input phase profile imparted by the SLM and the subset of the beam that is extracted by a spatial filter, due to intensity correlations that form between different parts of the beam. By sweeping over the SLM profile and the spatial filter, low noise can be realized at higher intensities. (\textbf{b}) In the experiment, we sweep over both. All noise levels shown are given relative to the noise level of the original source after linear attenuation (LA) to the same intensity. Input control substantially widens the space of noise behaviors compared to the same experiment performed without SLM variation (orange line). (\textbf{c}) Comparison of the experiment to simulations performed with our theoretical framework, showing good agreement with the experiment. The simulations confirm that some of the shaped initial conditions can bring the noise of these random quantities down to or below the standard quantum limit (SQL).}
    \label{fig:3}
\end{figure*}

\subsection*{Optimizing noise via input shaping and programmable spatial filtering}

We now demonstrate further optimization of the combined intensity and intensity noise properties by spatial filtering of the output beam. With this method, we find subsets of the beam with even higher power but with low noise, outperforming a single ``pixel'' on the Fano factor (noise divided by intensity). We isolate subsets of the output beam via an arbitrarily programmable spatial filter, realized by activating several of the mirrors of the DMD (Fig.~\ref{fig:3}a). We additionally shape the input phase profile via an SLM, which we will show is essential for realizing optimal noise performance. We approach the problem of sweeping the space of spatial filters by randomly activating subsets of DMD pixels, thereby routing some portion of the beam to the detector.

Results from this experiment are shown in Fig.~\ref{fig:3}b, where we show a histogram of the resulting noise as a function of the measured output transmission associated with a spatial filter in question. For low transmission, the sampled noise levels converge toward the LA limit due to the dominance of the noise floor. For high transmission, the behavior also converges toward the LA limit; this occurs since, in the limit of sampling the whole beam, there is no attenuation. Between these two extremes, the samples span a wide range of noise levels, from 5 dB of noise addition to more than 10 dB of noise reduction. The most intense noise-reduced states are observed at a transmission around $\sim0.5\%$. Moreover, the histogram colorbar reveals that most of the samples are concentrated around $\pm 1$ dB of the LA limit, while both high and low noise events are rarer. 

This increased range of noise behaviors is possible due to correlations between different pixels, which we have experimentally verified (see SI). More explicitly, the variance in the sum of many pixel intensities is given by $\sum_i (\Delta n_i)^2 + 2\sum_{i < j}\mathrm{cov}(n_i, n_j)$; thus, sampling more pixels can lead to destructive (or constructive) interference of noise across different parts of the beam. To determine the role of input shaping for achieving these behaviors, we repeat the random output sampling experiment in the absence of wavefront shaping. The resulting data (see Fig.~\ref{fig:3}b, orange curve) shows that without wavefront shaping, the sampled noise behaviors are much less diverse (around $\pm 1$ dB compared to LA). This leads us to conclude that the more complex nonlinear dynamics enabled by wavefront shaping also generate a wider diversity of correlations between fiber degrees of freedom, and thus the sampled pixels. 

We reproduce these experimental results in our simulation framework by randomly sampling initial conditions as well as the output pixels (Fig.~\ref{fig:3}c). The simulations generally agree well with the experiment, reproducing the overall shape of the envelope of sampled states, as well as the transmission and noise reduction levels of the lowest noise states. This comparison, in combination with a fit to the shot-noise level based on experimental data (see SI), indicates that the low-noise states observed in the experiment (up to transmission $\sim0.5\%$) are at a quantum noise level, i.e., within a few decibels of the shot noise level. This is particularly remarkable, given that the initial beam has high excess noise from laser amplification. The excess noise of the beam at 260 mW is estimated to be 30 dB above the shot noise level (see SI). In particular, higher levels of input noise make it difficult to produce light with quantum levels of noise, even with strong attenuation. However, the combination of programmable wavefront shaping, programmable output filtering, and nonlinearity enables very strong noise reduction. This observation suggests a path for using spatial filtering to develop conventionally amplified light sources that operate near quantum noise. The simulations also reveal that this procedure allows, in principle, for noise reduction below shot noise level, dependent on input power and linear loss --- here the maximum squeezing is $\sim 0.5$ dB. We note here that the only free parameters that enter our simulation are the noise floor, the excess noise in the incoming beam, the loss at the DMD, and an estimation of the power scale.

These results demonstrate that input shaping of a multimode nonlinear system can augment the range of possible noise behaviors at the output. In the absence of shaping, the correlations that build up inside the system tend to result in excess noise at the output, which is often undesirable. For certain special choices of either the input phase profile or the output spatial filter, these nonlinearly generated correlations are beneficial and allow the system to fluctuate less than what could be achieved by linear attenuation. In this case, a programmed nonlinearity acts as a resource to generate correlations that suppress noise.

\subsection*{Power dependence of spatial noise}

\begin{figure*}[t!]
    \centering
    \includegraphics{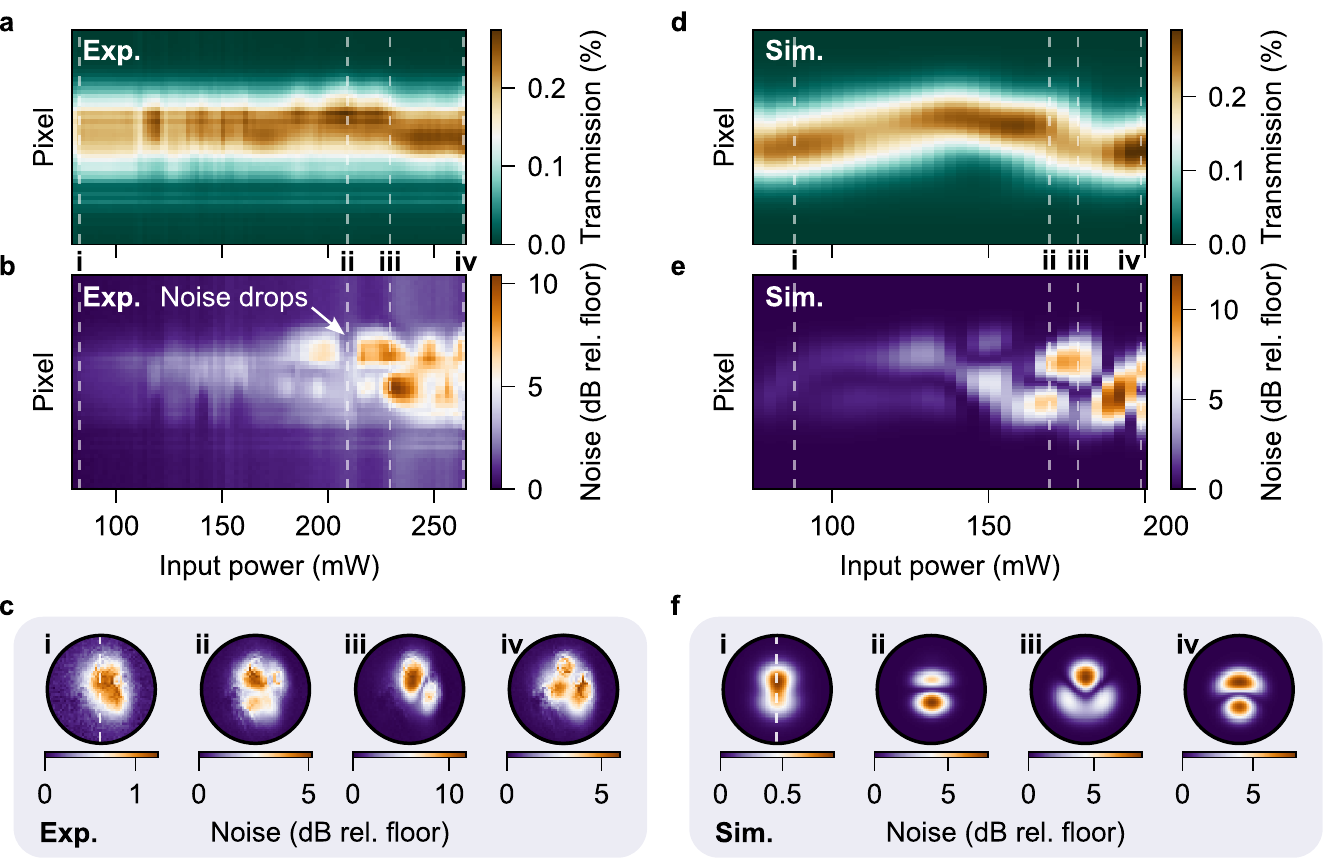}
    \caption{\textbf{Power dependent spatial noise dynamics.} (\textbf{a}) Experimental transmission of light across a line-cut across the beam (shown by the dashed line in c-i), as a function of the average input power. (\textbf{b}) Measured noise for the same input powers and pixels shown in (a). (\textbf{c}) Noise distribution of the whole beam for several selected input powers. (\textbf{d}-\textbf{f}) Same as (a-c) for the simulation model which best reproduces the experimental behavior.}
    \label{fig:4}
\end{figure*}

Having established that wavefront shaping and spatial filtering can strongly influence quantum noise dynamics, we now elucidate the physical mechanisms that lead to intensity noise build-up in the absence of wavefront shaping, i.e., due to nonlinear dynamics alone. To accomplish this, we experimentally probed the power dependence of intensity noise by exciting the system with a simple initial condition (a focused spot which is slightly off-center at the input, see Fig.~\ref{fig:schematic}a). We then measured the output intensity and noise images of the beam as a function of the input power.

The variations of the transmission and noise can be visualized through a power-dependent line-cut of pixels across the beam (Fig.~\ref{fig:4}a-b). Nonlinearity causes minor variations to the transmission behavior across powers (consisting mostly of side-to-side motion of the beam and distortions) coming from the self-imaging properties of a gradient-index fiber for a Gaussian input mode \cite{agrawal_invite_2019}. On the other hand, nonlinearity induces much more discernible changes in the noise behavior (Fig.~\ref{fig:4}b). The noise images taken at various powers show the formation of multi-lobed structures at higher powers (Fig.~\ref{fig:4}c). The full dataset of intensity and noise images at all powers is provided in the SI.

To explain these behaviors, we developed a simulation model based on the full spatio-temporal dynamics in the experiment (Fig.~\ref{fig:4}d-f). The classical simulations indicate that the input pulse populates a small number of spatial modes (with the fundamental mode being preferentially populated). During propagation through the fiber, the modal population exchange is relatively weak (only a few percent at the highest powers). Thus, the power-dependent behavior of the output beam is primarily determined by nonlinear changes to the temporal overlap between different spatial modes, induced by self- and cross-phase modulation. 

We also developed simulations of the noise in this model, which show several elements of strong agreement with our experimental data. First, the simulation correctly captures the trend that the multimode nonlinearity generally adds more noise with increasing input power. Secondly, both the experiment and simulation show several regions of behavior where the noise can be interpreted in terms of power-dependent changes in transmission. For example, there are regions where sharply increasing transmission is accompanied by a noise peak; there are also regions where a ``flat'' transmission is accompanied by a sizable drop in the noise level. Finally, the simulations readily reproduce the types of multi-lobed noise structures that were experimentally observed, with some deviations which we attribute to experimental uncertainties related to the exact pulse profile and the fiber itself (Fig.~\ref{fig:4}f).

\begin{figure}[t!]
    \centering
    \includegraphics{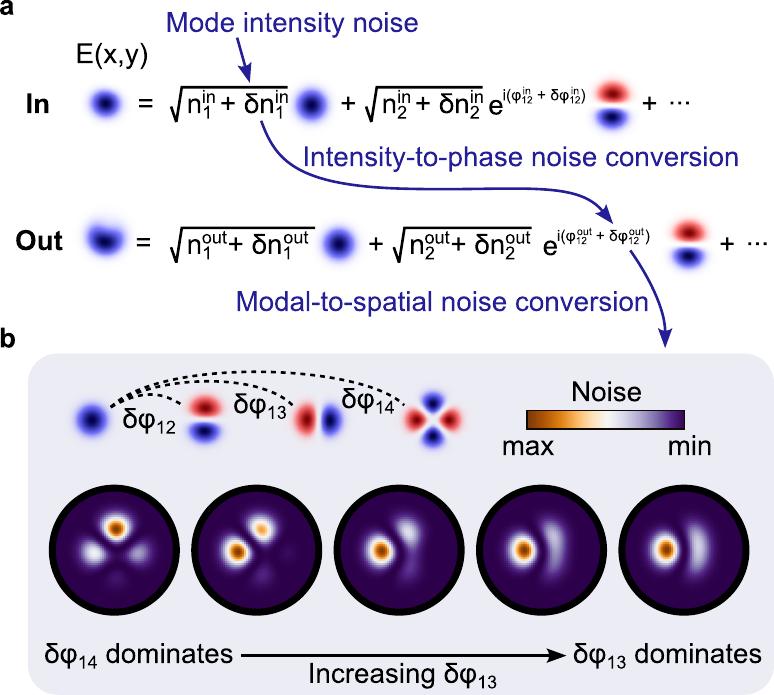}
    \caption{\textbf{Mechanisms of noise buildup in spatially multimode nonlinear optics.} (\textbf{a}) The noisy field in the fiber can be expanded into spatial modes with amplitudes and phases, each of which has fluctuations. Due to the nonlinear interactions, intensity fluctuations at the input couple to intermodal phase fluctuations at the output. When these modes interfere to form an image, intensity noise patterns are generated that bear the signature of the dominant noisy mode interference. (\textbf{b}) Example of how changing the distribution of phase noise between the fundamental mode and the new few higher-order modes creates different intensity noise patterns. In this example, $\delta\varphi_{12}$ and $\delta\varphi_{14}$ are fixed, while $\delta\varphi_{13}$ increases from left to right.}
    \label{fig:5}
\end{figure}

Our theoretical model reveals that these multi-lobed spatial noise patterns arise from the fluctuations in the relative phase between spatial modes, leading to corresponding fluctuations in the interference between the fields in different spatial modes. We developed a simplified model of the full nonlinear dynamics which traces noise through the system in several stages (Fig.~\ref{fig:5}a). In this model, we consider the noisy electric field in the fiber to be expanded as $E(x,y) = \sum_m \sqrt{n_m + \delta n_m} e^{\mathrm{i}(\varphi_m + \delta\varphi_m)} \phi_m(x,y)$, where $n_m$ and $\varphi_m$ are, respectively, the number of photons and phase of mode $m$ which has profile $\phi_m(x,y)$ representing the electric field profile of mode $m$. Moreover, $\delta n_m$ and $\delta \varphi_m$ are the fluctuations in those quantities. The theory and simulations reveal a dominant mechanism of excess noise generation in our experiments: the nonlinearity-induced conversion of power fluctuations at the input into phase fluctuations at the output. Kerr nonlinearity in the fiber causes strong self- and cross-phase modulation of pulses occupying the different spatial modes. Even minor changes to the input power can cause substantial changes in these dynamics, resulting in high levels of noise in the phases between different modes. Then, the overlap between different beating mode pairs generates various spatial noise patterns (Fig.~\ref{fig:5}b). For example, a reduced model which incorporates phase noise only $\delta\varphi_{ml}$ between the fundamental mode ($m=1$) and the next few modes ($l=\{2,3,4\}$) produces spatial noise profiles that strongly resemble the experimentally observed ones. Thus, these spatial patterns of noise provide information about which mode pairs' temporal interference is most (or least) sensitive to noise at the input. 

This perspective explains the physical behavior that occurs at some of the powers that generate the highest overall noise levels (Fig.~\ref{fig:4}). At some of these points, excess input noise can be amplified into phase fluctuations between the two most populated modes (fundamental and dipole). In this case, a two-lobed noise pattern is generated (e.g., Fig.~\ref{fig:4}c-iii, Fig.~\ref{fig:4}f-ii). It also explains one of the most striking experimental observations: a narrow range of input powers where the noise level suddenly drops, while the noise pattern simultaneously transitions from two-lobed to multi-lobed (Fig.~\ref{fig:4}c-ii). For this particular power, the power-dependent changes to the beam are quite stationary (in contrast to many other powers, where changing the power shifts the beam substantially). Thus, we identify this power as a point where the relative phase between the most populated modes has very low noise. As a result, fluctuations involving higher-order and relatively unpopulated modes dominate, forming a multi-lobed pattern.  

\subsection*{Outlook}

In summary, we have demonstrated that the noise properties of light at the output of a nonlinear and spatiotemporally multimode system can be strongly influenced by controlling the initial conditions with wavefront shaping. The general concept was demonstrated based on a feedback and optimization approach on the intensity noise of light emerging from a spatially multimode fiber. Additionally, we showed that a generic mechanism of noise generation in spatially multimode systems is via the coupling of incident intensity noise (in our case, excess noise arising from amplified quantum noise) to noise in the relative phase between spatial modes, leading to strong noise in the spatially-dependent intensity. This noise generating effect can increase strongly with power and is expected to be present in general multimode nonlinear systems operated at high powers, which are needed in applications such as white-light generation \cite{wright_controllable_2015}, high-power fiber lasers \cite{richardson_high_2010,wright_spatiotemporal_2017}, communication systems based on space-division-multiplexing \cite{richardson_space-division_2013,puttnam_space-division_2021}, and levitated optomechanics \cite{gonzalez-ballestero_suppressing_2023}.

Our results provide a new perspective on the design of high-performance light sources. By exploiting the programmability of multimode dynamics, one can create systems with desired average properties (e.g., spatial, temporal/spectral, polarization), while also having low noise in some quantity of interest (e.g., intensity, phase, jitter). Importantly, these quantities can be designed to have fluctuations down to quantum levels, even in the presence of amplifiers, which tend to generate noise. Application of the concept demonstrated here --- using programmable nonlinearities to shape quantum noise --- to sources already operating closer to the shot-noise level should enable the controlled generation of squeezed states of light by complex nonlinear systems. It may even become possible to exploit programmability to create arbitrary Gaussian states, including entangled states such as cluster states. Finally, these same wavefront-shaping ideas, applied directly in the quantum domain, could protect fragile quantum states from noise-adding processes such as amplification or random scattering, by steering light into modes that are minimally affected.

\subsection*{Acknowledgements}

We acknowledge useful discussions with Nicholas Bender and Frank Wise.  J.S. acknowledges previous support of a Mathworks Fellowship, as well as previous support from a National Defense Science and Engineering Graduate (NDSEG) Fellowship (F-1730184536). This research was funded in part (M.H.) by the Austrian Science Fund (FWF) [10.55776/J4729].  This material is based upon work also supported in part by the U. S. Army Research Office through the Institute for Soldier Nanotechnologies at MIT, under Collaborative Agreement Number W911NF-23-2-0121. M.S. also acknowledges support of Parviz Tayebati. This publication was also supported in part by the DARPA Agreement number HO0011249049. The authors acknowledge the MIT SuperCloud and Lincoln Laboratory Supercomputing Center for providing high-performance computing resources that have contributed to the research results reported within this paper. N.R. acknowledges prior support from a Junior Fellowship from the Harvard Society of Fellows, as well as start-up funding from the School of Applied and Engineering Physics at Cornell University.

\bibliographystyle{apsrev4-2}

\begin{thebibliography}{57}%
\makeatletter
\providecommand \@ifxundefined [1]{%
 \@ifx{#1\undefined}
}%
\providecommand \@ifnum [1]{%
 \ifnum #1\expandafter \@firstoftwo
 \else \expandafter \@secondoftwo
 \fi
}%
\providecommand \@ifx [1]{%
 \ifx #1\expandafter \@firstoftwo
 \else \expandafter \@secondoftwo
 \fi
}%
\providecommand \natexlab [1]{#1}%
\providecommand \enquote  [1]{``#1''}%
\providecommand \bibnamefont  [1]{#1}%
\providecommand \bibfnamefont [1]{#1}%
\providecommand \citenamefont [1]{#1}%
\providecommand \href@noop [0]{\@secondoftwo}%
\providecommand \href [0]{\begingroup \@sanitize@url \@href}%
\providecommand \@href[1]{\@@startlink{#1}\@@href}%
\providecommand \@@href[1]{\endgroup#1\@@endlink}%
\providecommand \@sanitize@url [0]{\catcode `\\12\catcode `\$12\catcode `\&12\catcode `\#12\catcode `\^12\catcode `\_12\catcode `\%12\relax}%
\providecommand \@@startlink[1]{}%
\providecommand \@@endlink[0]{}%
\providecommand \url  [0]{\begingroup\@sanitize@url \@url }%
\providecommand \@url [1]{\endgroup\@href {#1}{\urlprefix }}%
\providecommand \urlprefix  [0]{URL }%
\providecommand \Eprint [0]{\href }%
\providecommand \doibase [0]{https://doi.org/}%
\providecommand \selectlanguage [0]{\@gobble}%
\providecommand \bibinfo  [0]{\@secondoftwo}%
\providecommand \bibfield  [0]{\@secondoftwo}%
\providecommand \translation [1]{[#1]}%
\providecommand \BibitemOpen [0]{}%
\providecommand \bibitemStop [0]{}%
\providecommand \bibitemNoStop [0]{.\EOS\space}%
\providecommand \EOS [0]{\spacefactor3000\relax}%
\providecommand \BibitemShut  [1]{\csname bibitem#1\endcsname}%
\let\auto@bib@innerbib\@empty
\bibitem [{\citenamefont {Hecht}(2010)}]{hecht_how_2010}%
  \BibitemOpen
  \bibfield  {author} {\bibinfo {author} {\bibfnamefont {J.}~\bibnamefont {Hecht}},\ }\href {https://www.optica-opn.org/home/articles/volume_21/issue_11/features/how_the_laser_launched_nonlinear_optics/} {\bibfield  {journal} {\bibinfo  {journal} {Optics \& Photonics News}\ }\textbf {\bibinfo {volume} {21}} (\bibinfo {year} {2010})}\BibitemShut {NoStop}%
\bibitem [{\citenamefont {Franken}\ \emph {et~al.}(1961)\citenamefont {Franken}, \citenamefont {Hill}, \citenamefont {Peters},\ and\ \citenamefont {Weinreich}}]{franken_generation_1961}%
  \BibitemOpen
  \bibfield  {author} {\bibinfo {author} {\bibfnamefont {P.~A.}\ \bibnamefont {Franken}}, \bibinfo {author} {\bibfnamefont {A.~E.}\ \bibnamefont {Hill}}, \bibinfo {author} {\bibfnamefont {C.~W.}\ \bibnamefont {Peters}},\ and\ \bibinfo {author} {\bibfnamefont {G.}~\bibnamefont {Weinreich}},\ }\href {https://doi.org/10.1103/PhysRevLett.7.118} {\bibfield  {journal} {\bibinfo  {journal} {Physical Review Letters}\ }\textbf {\bibinfo {volume} {7}},\ \bibinfo {pages} {118} (\bibinfo {year} {1961})}\BibitemShut {NoStop}%
\bibitem [{\citenamefont {Keller}(2003)}]{keller_recent_2003}%
  \BibitemOpen
  \bibfield  {author} {\bibinfo {author} {\bibfnamefont {U.}~\bibnamefont {Keller}},\ }\href {https://doi.org/10.1038/nature01938} {\bibfield  {journal} {\bibinfo  {journal} {Nature}\ }\textbf {\bibinfo {volume} {424}},\ \bibinfo {pages} {831} (\bibinfo {year} {2003})}\BibitemShut {NoStop}%
\bibitem [{\citenamefont {Denk}\ \emph {et~al.}(1990)\citenamefont {Denk}, \citenamefont {Strickler},\ and\ \citenamefont {Webb}}]{denk_two-photon_1990}%
  \BibitemOpen
  \bibfield  {author} {\bibinfo {author} {\bibfnamefont {W.}~\bibnamefont {Denk}}, \bibinfo {author} {\bibfnamefont {J.~H.}\ \bibnamefont {Strickler}},\ and\ \bibinfo {author} {\bibfnamefont {W.~W.}\ \bibnamefont {Webb}},\ }\href {https://doi.org/10.1126/science.2321027} {\bibfield  {journal} {\bibinfo  {journal} {Science}\ }\textbf {\bibinfo {volume} {248}},\ \bibinfo {pages} {73} (\bibinfo {year} {1990})}\BibitemShut {NoStop}%
\bibitem [{\citenamefont {Hell}\ and\ \citenamefont {Wichmann}(1994)}]{hell_breaking_1994}%
  \BibitemOpen
  \bibfield  {author} {\bibinfo {author} {\bibfnamefont {S.~W.}\ \bibnamefont {Hell}}\ and\ \bibinfo {author} {\bibfnamefont {J.}~\bibnamefont {Wichmann}},\ }\href {https://doi.org/10.1364/OL.19.000780} {\bibfield  {journal} {\bibinfo  {journal} {Optics Letters}\ }\textbf {\bibinfo {volume} {19}},\ \bibinfo {pages} {780} (\bibinfo {year} {1994})}\BibitemShut {NoStop}%
\bibitem [{\citenamefont {Freudiger}\ \emph {et~al.}(2008)\citenamefont {Freudiger}, \citenamefont {Min}, \citenamefont {Saar}, \citenamefont {Lu}, \citenamefont {Holtom}, \citenamefont {He}, \citenamefont {Tsai}, \citenamefont {Kang},\ and\ \citenamefont {Xie}}]{freudiger_label-free_2008}%
  \BibitemOpen
  \bibfield  {author} {\bibinfo {author} {\bibfnamefont {C.~W.}\ \bibnamefont {Freudiger}}, \bibinfo {author} {\bibfnamefont {W.}~\bibnamefont {Min}}, \bibinfo {author} {\bibfnamefont {B.~G.}\ \bibnamefont {Saar}}, \bibinfo {author} {\bibfnamefont {S.}~\bibnamefont {Lu}}, \bibinfo {author} {\bibfnamefont {G.~R.}\ \bibnamefont {Holtom}}, \bibinfo {author} {\bibfnamefont {C.}~\bibnamefont {He}}, \bibinfo {author} {\bibfnamefont {J.~C.}\ \bibnamefont {Tsai}}, \bibinfo {author} {\bibfnamefont {J.~X.}\ \bibnamefont {Kang}},\ and\ \bibinfo {author} {\bibfnamefont {X.~S.}\ \bibnamefont {Xie}},\ }\href {https://doi.org/10.1126/science.1165758} {\bibfield  {journal} {\bibinfo  {journal} {Science}\ }\textbf {\bibinfo {volume} {322}},\ \bibinfo {pages} {1857} (\bibinfo {year} {2008})}\BibitemShut {NoStop}%
\bibitem [{\citenamefont {Wright}\ \emph {et~al.}(2022{\natexlab{a}})\citenamefont {Wright}, \citenamefont {Wu}, \citenamefont {Christodoulides},\ and\ \citenamefont {Wise}}]{wright_physics_2022}%
  \BibitemOpen
  \bibfield  {author} {\bibinfo {author} {\bibfnamefont {L.~G.}\ \bibnamefont {Wright}}, \bibinfo {author} {\bibfnamefont {F.~O.}\ \bibnamefont {Wu}}, \bibinfo {author} {\bibfnamefont {D.~N.}\ \bibnamefont {Christodoulides}},\ and\ \bibinfo {author} {\bibfnamefont {F.~W.}\ \bibnamefont {Wise}},\ }\href {https://doi.org/10.1038/s41567-022-01691-z} {\bibfield  {journal} {\bibinfo  {journal} {Nature Physics}\ }\textbf {\bibinfo {volume} {18}},\ \bibinfo {pages} {1018} (\bibinfo {year} {2022}{\natexlab{a}})}\BibitemShut {NoStop}%
\bibitem [{\citenamefont {Lu}\ \emph {et~al.}(2023)\citenamefont {Lu}, \citenamefont {Liscidini}, \citenamefont {Gaeta}, \citenamefont {Weiner},\ and\ \citenamefont {Lukens}}]{lu_frequency-bin_2023}%
  \BibitemOpen
  \bibfield  {author} {\bibinfo {author} {\bibfnamefont {H.-H.}\ \bibnamefont {Lu}}, \bibinfo {author} {\bibfnamefont {M.}~\bibnamefont {Liscidini}}, \bibinfo {author} {\bibfnamefont {A.~L.}\ \bibnamefont {Gaeta}}, \bibinfo {author} {\bibfnamefont {A.~M.}\ \bibnamefont {Weiner}},\ and\ \bibinfo {author} {\bibfnamefont {J.~M.}\ \bibnamefont {Lukens}},\ }\href {https://doi.org/10.1364/OPTICA.506096} {\bibfield  {journal} {\bibinfo  {journal} {Optica}\ }\textbf {\bibinfo {volume} {10}},\ \bibinfo {pages} {1655} (\bibinfo {year} {2023})}\BibitemShut {NoStop}%
\bibitem [{\citenamefont {Puttnam}\ \emph {et~al.}(2021)\citenamefont {Puttnam}, \citenamefont {Rademacher},\ and\ \citenamefont {Luís}}]{puttnam_space-division_2021}%
  \BibitemOpen
  \bibfield  {author} {\bibinfo {author} {\bibfnamefont {B.~J.}\ \bibnamefont {Puttnam}}, \bibinfo {author} {\bibfnamefont {G.}~\bibnamefont {Rademacher}},\ and\ \bibinfo {author} {\bibfnamefont {R.~S.}\ \bibnamefont {Luís}},\ }\href {https://doi.org/10.1364/OPTICA.427631} {\bibfield  {journal} {\bibinfo  {journal} {Optica}\ }\textbf {\bibinfo {volume} {8}},\ \bibinfo {pages} {1186} (\bibinfo {year} {2021})}\BibitemShut {NoStop}%
\bibitem [{\citenamefont {Wang}\ \emph {et~al.}(2022)\citenamefont {Wang}, \citenamefont {Liu}, \citenamefont {Li}, \citenamefont {Zhao}, \citenamefont {Du},\ and\ \citenamefont {Zhu}}]{wang_orbital_2022}%
  \BibitemOpen
  \bibfield  {author} {\bibinfo {author} {\bibfnamefont {J.}~\bibnamefont {Wang}}, \bibinfo {author} {\bibfnamefont {J.}~\bibnamefont {Liu}}, \bibinfo {author} {\bibfnamefont {S.}~\bibnamefont {Li}}, \bibinfo {author} {\bibfnamefont {Y.}~\bibnamefont {Zhao}}, \bibinfo {author} {\bibfnamefont {J.}~\bibnamefont {Du}},\ and\ \bibinfo {author} {\bibfnamefont {L.}~\bibnamefont {Zhu}},\ }\href {https://doi.org/10.1515/nanoph-2021-0527} {\bibfield  {journal} {\bibinfo  {journal} {Nanophotonics}\ }\textbf {\bibinfo {volume} {11}},\ \bibinfo {pages} {645} (\bibinfo {year} {2022})}\BibitemShut {NoStop}%
\bibitem [{\citenamefont {Wright}\ \emph {et~al.}(2022{\natexlab{b}})\citenamefont {Wright}, \citenamefont {Renninger}, \citenamefont {Christodoulides},\ and\ \citenamefont {Wise}}]{wright_nonlinear_2022}%
  \BibitemOpen
  \bibfield  {author} {\bibinfo {author} {\bibfnamefont {L.~G.}\ \bibnamefont {Wright}}, \bibinfo {author} {\bibfnamefont {W.~H.}\ \bibnamefont {Renninger}}, \bibinfo {author} {\bibfnamefont {D.~N.}\ \bibnamefont {Christodoulides}},\ and\ \bibinfo {author} {\bibfnamefont {F.~W.}\ \bibnamefont {Wise}},\ }\href {https://doi.org/10.1364/OPTICA.461981} {\bibfield  {journal} {\bibinfo  {journal} {Optica}\ }\textbf {\bibinfo {volume} {9}},\ \bibinfo {pages} {824} (\bibinfo {year} {2022}{\natexlab{b}})}\BibitemShut {NoStop}%
\bibitem [{\citenamefont {Vitali}\ \emph {et~al.}(2025)\citenamefont {Vitali}, \citenamefont {Bucio}, \citenamefont {Liu}, \citenamefont {Haines}, \citenamefont {Naik}, \citenamefont {Guasoni}, \citenamefont {Gardes}, \citenamefont {Pavesi}, \citenamefont {Cristiani}, \citenamefont {Lacava},\ and\ \citenamefont {Petropoulos}}]{vitali_nonlinear_2025}%
  \BibitemOpen
  \bibfield  {author} {\bibinfo {author} {\bibfnamefont {V.}~\bibnamefont {Vitali}}, \bibinfo {author} {\bibfnamefont {T.~D.}\ \bibnamefont {Bucio}}, \bibinfo {author} {\bibfnamefont {H.}~\bibnamefont {Liu}}, \bibinfo {author} {\bibfnamefont {J.}~\bibnamefont {Haines}}, \bibinfo {author} {\bibfnamefont {P.~U.}\ \bibnamefont {Naik}}, \bibinfo {author} {\bibfnamefont {M.}~\bibnamefont {Guasoni}}, \bibinfo {author} {\bibfnamefont {F.}~\bibnamefont {Gardes}}, \bibinfo {author} {\bibfnamefont {L.}~\bibnamefont {Pavesi}}, \bibinfo {author} {\bibfnamefont {I.}~\bibnamefont {Cristiani}}, \bibinfo {author} {\bibfnamefont {C.}~\bibnamefont {Lacava}},\ and\ \bibinfo {author} {\bibfnamefont {P.}~\bibnamefont {Petropoulos}},\ }\href {https://doi.org/10.1515/nanoph-2025-0105} {\bibfield  {journal} {\bibinfo  {journal} {Nanophotonics}\ }\textbf {\bibinfo {volume} {14}},\ \bibinfo {pages} {2507} (\bibinfo {year} {2025})}\BibitemShut {NoStop}%
\bibitem [{\citenamefont {Chen}\ \emph {et~al.}(2023{\natexlab{a}})\citenamefont {Chen}, \citenamefont {Wisal}, \citenamefont {Eliezer}, \citenamefont {Stone},\ and\ \citenamefont {Cao}}]{chen_suppressing_2023}%
  \BibitemOpen
  \bibfield  {author} {\bibinfo {author} {\bibfnamefont {C.-W.}\ \bibnamefont {Chen}}, \bibinfo {author} {\bibfnamefont {K.}~\bibnamefont {Wisal}}, \bibinfo {author} {\bibfnamefont {Y.}~\bibnamefont {Eliezer}}, \bibinfo {author} {\bibfnamefont {A.~D.}\ \bibnamefont {Stone}},\ and\ \bibinfo {author} {\bibfnamefont {H.}~\bibnamefont {Cao}},\ }\href {https://doi.org/10.1073/pnas.2217735120} {\bibfield  {journal} {\bibinfo  {journal} {Proceedings of the National Academy of Sciences}\ }\textbf {\bibinfo {volume} {120}},\ \bibinfo {pages} {e2217735120} (\bibinfo {year} {2023}{\natexlab{a}})}\BibitemShut {NoStop}%
\bibitem [{\citenamefont {Rothe}\ \emph {et~al.}(2025)\citenamefont {Rothe}, \citenamefont {CHen}, \citenamefont {Ahmadi}, \citenamefont {Wisal}, \citenamefont {Ercan}, \citenamefont {Lee}, \citenamefont {VIgne}, \citenamefont {Stone},\ and\ \citenamefont {Cao}}]{rothe_wavefront_2025}%
  \BibitemOpen
  \bibfield  {author} {\bibinfo {author} {\bibfnamefont {S.}~\bibnamefont {Rothe}}, \bibinfo {author} {\bibfnamefont {C.-W.}\ \bibnamefont {CHen}}, \bibinfo {author} {\bibfnamefont {P.}~\bibnamefont {Ahmadi}}, \bibinfo {author} {\bibfnamefont {K.}~\bibnamefont {Wisal}}, \bibinfo {author} {\bibfnamefont {M.}~\bibnamefont {Ercan}}, \bibinfo {author} {\bibfnamefont {K.}~\bibnamefont {Lee}}, \bibinfo {author} {\bibfnamefont {N.}~\bibnamefont {VIgne}}, \bibinfo {author} {\bibfnamefont {A.~D.}\ \bibnamefont {Stone}},\ and\ \bibinfo {author} {\bibfnamefont {H.}~\bibnamefont {Cao}},\ }\href {https://doi.org/10.48550/arXiv.2504.06423} {\bibinfo {title} {Wavefront shaping enables high-power multimode fiber amplifier with output control}} (\bibinfo {year} {2025}),\ \bibinfo {note} {arXiv:2504.06423 [physics]}\BibitemShut {NoStop}%
\bibitem [{\citenamefont {Richardson}\ \emph {et~al.}(2013)\citenamefont {Richardson}, \citenamefont {Fini},\ and\ \citenamefont {Nelson}}]{richardson_space-division_2013}%
  \BibitemOpen
  \bibfield  {author} {\bibinfo {author} {\bibfnamefont {D.~J.}\ \bibnamefont {Richardson}}, \bibinfo {author} {\bibfnamefont {J.~M.}\ \bibnamefont {Fini}},\ and\ \bibinfo {author} {\bibfnamefont {L.~E.}\ \bibnamefont {Nelson}},\ }\href {https://doi.org/10.1038/nphoton.2013.94} {\bibfield  {journal} {\bibinfo  {journal} {Nature Photonics}\ }\textbf {\bibinfo {volume} {7}},\ \bibinfo {pages} {354} (\bibinfo {year} {2013})}\BibitemShut {NoStop}%
\bibitem [{\citenamefont {Haus}(2000)}]{haus_mode-locking_2000}%
  \BibitemOpen
  \bibfield  {author} {\bibinfo {author} {\bibfnamefont {H.}~\bibnamefont {Haus}},\ }\href {https://doi.org/10.1109/2944.902165} {\bibfield  {journal} {\bibinfo  {journal} {IEEE Journal of Selected Topics in Quantum Electronics}\ }\textbf {\bibinfo {volume} {6}},\ \bibinfo {pages} {1173} (\bibinfo {year} {2000})}\BibitemShut {NoStop}%
\bibitem [{\citenamefont {Wu}\ \emph {et~al.}(2019)\citenamefont {Wu}, \citenamefont {Hassan},\ and\ \citenamefont {Christodoulides}}]{wu_thermodynamic_2019}%
  \BibitemOpen
  \bibfield  {author} {\bibinfo {author} {\bibfnamefont {F.~O.}\ \bibnamefont {Wu}}, \bibinfo {author} {\bibfnamefont {A.~U.}\ \bibnamefont {Hassan}},\ and\ \bibinfo {author} {\bibfnamefont {D.~N.}\ \bibnamefont {Christodoulides}},\ }\href {https://doi.org/10.1038/s41566-019-0501-8} {\bibfield  {journal} {\bibinfo  {journal} {Nature Photonics}\ }\textbf {\bibinfo {volume} {13}},\ \bibinfo {pages} {776} (\bibinfo {year} {2019})}\BibitemShut {NoStop}%
\bibitem [{\citenamefont {Pourbeyram}\ \emph {et~al.}(2022)\citenamefont {Pourbeyram}, \citenamefont {Sidorenko}, \citenamefont {Wu}, \citenamefont {Bender}, \citenamefont {Wright}, \citenamefont {Christodoulides},\ and\ \citenamefont {Wise}}]{pourbeyram_direct_2022}%
  \BibitemOpen
  \bibfield  {author} {\bibinfo {author} {\bibfnamefont {H.}~\bibnamefont {Pourbeyram}}, \bibinfo {author} {\bibfnamefont {P.}~\bibnamefont {Sidorenko}}, \bibinfo {author} {\bibfnamefont {F.~O.}\ \bibnamefont {Wu}}, \bibinfo {author} {\bibfnamefont {N.}~\bibnamefont {Bender}}, \bibinfo {author} {\bibfnamefont {L.}~\bibnamefont {Wright}}, \bibinfo {author} {\bibfnamefont {D.~N.}\ \bibnamefont {Christodoulides}},\ and\ \bibinfo {author} {\bibfnamefont {F.}~\bibnamefont {Wise}},\ }\href {https://doi.org/10.1038/s41567-022-01579-y} {\bibfield  {journal} {\bibinfo  {journal} {Nature Physics}\ }\textbf {\bibinfo {volume} {18}},\ \bibinfo {pages} {685} (\bibinfo {year} {2022})}\BibitemShut {NoStop}%
\bibitem [{\citenamefont {Wright}\ \emph {et~al.}(2017)\citenamefont {Wright}, \citenamefont {Christodoulides},\ and\ \citenamefont {Wise}}]{wright_spatiotemporal_2017}%
  \BibitemOpen
  \bibfield  {author} {\bibinfo {author} {\bibfnamefont {L.~G.}\ \bibnamefont {Wright}}, \bibinfo {author} {\bibfnamefont {D.~N.}\ \bibnamefont {Christodoulides}},\ and\ \bibinfo {author} {\bibfnamefont {F.~W.}\ \bibnamefont {Wise}},\ }\href {https://doi.org/10.1126/science.aao0831} {\bibfield  {journal} {\bibinfo  {journal} {Science}\ }\textbf {\bibinfo {volume} {358}},\ \bibinfo {pages} {94} (\bibinfo {year} {2017})}\BibitemShut {NoStop}%
\bibitem [{\citenamefont {Wright}\ \emph {et~al.}(2022{\natexlab{c}})\citenamefont {Wright}, \citenamefont {Onodera}, \citenamefont {Stein}, \citenamefont {Wang}, \citenamefont {Schachter}, \citenamefont {Hu},\ and\ \citenamefont {McMahon}}]{wright_deep_2022}%
  \BibitemOpen
  \bibfield  {author} {\bibinfo {author} {\bibfnamefont {L.~G.}\ \bibnamefont {Wright}}, \bibinfo {author} {\bibfnamefont {T.}~\bibnamefont {Onodera}}, \bibinfo {author} {\bibfnamefont {M.~M.}\ \bibnamefont {Stein}}, \bibinfo {author} {\bibfnamefont {T.}~\bibnamefont {Wang}}, \bibinfo {author} {\bibfnamefont {D.~T.}\ \bibnamefont {Schachter}}, \bibinfo {author} {\bibfnamefont {Z.}~\bibnamefont {Hu}},\ and\ \bibinfo {author} {\bibfnamefont {P.~L.}\ \bibnamefont {McMahon}},\ }\href {https://doi.org/10.1038/s41586-021-04223-6} {\bibfield  {journal} {\bibinfo  {journal} {Nature}\ }\textbf {\bibinfo {volume} {601}},\ \bibinfo {pages} {549} (\bibinfo {year} {2022}{\natexlab{c}})}\BibitemShut {NoStop}%
\bibitem [{\citenamefont {Momeni}\ \emph {et~al.}(2023)\citenamefont {Momeni}, \citenamefont {Rahmani}, \citenamefont {Malléjac}, \citenamefont {Del~Hougne},\ and\ \citenamefont {Fleury}}]{momeni_backpropagation-free_2023}%
  \BibitemOpen
  \bibfield  {author} {\bibinfo {author} {\bibfnamefont {A.}~\bibnamefont {Momeni}}, \bibinfo {author} {\bibfnamefont {B.}~\bibnamefont {Rahmani}}, \bibinfo {author} {\bibfnamefont {M.}~\bibnamefont {Malléjac}}, \bibinfo {author} {\bibfnamefont {P.}~\bibnamefont {Del~Hougne}},\ and\ \bibinfo {author} {\bibfnamefont {R.}~\bibnamefont {Fleury}},\ }\href {https://doi.org/10.1126/science.adi8474} {\bibfield  {journal} {\bibinfo  {journal} {Science}\ }\textbf {\bibinfo {volume} {382}},\ \bibinfo {pages} {1297} (\bibinfo {year} {2023})}\BibitemShut {NoStop}%
\bibitem [{\citenamefont {Guidry}\ \emph {et~al.}(2022)\citenamefont {Guidry}, \citenamefont {Lukin}, \citenamefont {Yang}, \citenamefont {Trivedi},\ and\ \citenamefont {Vučković}}]{guidry_quantum_2022}%
  \BibitemOpen
  \bibfield  {author} {\bibinfo {author} {\bibfnamefont {M.~A.}\ \bibnamefont {Guidry}}, \bibinfo {author} {\bibfnamefont {D.~M.}\ \bibnamefont {Lukin}}, \bibinfo {author} {\bibfnamefont {K.~Y.}\ \bibnamefont {Yang}}, \bibinfo {author} {\bibfnamefont {R.}~\bibnamefont {Trivedi}},\ and\ \bibinfo {author} {\bibfnamefont {J.}~\bibnamefont {Vučković}},\ }\href {https://doi.org/10.1038/s41566-021-00901-z} {\bibfield  {journal} {\bibinfo  {journal} {Nature Photonics}\ }\textbf {\bibinfo {volume} {16}},\ \bibinfo {pages} {52} (\bibinfo {year} {2022})},\ \bibinfo {note} {publisher: Springer Science and Business Media LLC}\BibitemShut {NoStop}%
\bibitem [{\citenamefont {Presutti}\ \emph {et~al.}(2024)\citenamefont {Presutti}, \citenamefont {Wright}, \citenamefont {Ma}, \citenamefont {Wang}, \citenamefont {Malia}, \citenamefont {Onodera},\ and\ \citenamefont {McMahon}}]{presutti_highly_2024}%
  \BibitemOpen
  \bibfield  {author} {\bibinfo {author} {\bibfnamefont {F.}~\bibnamefont {Presutti}}, \bibinfo {author} {\bibfnamefont {L.~G.}\ \bibnamefont {Wright}}, \bibinfo {author} {\bibfnamefont {S.-Y.}\ \bibnamefont {Ma}}, \bibinfo {author} {\bibfnamefont {T.}~\bibnamefont {Wang}}, \bibinfo {author} {\bibfnamefont {B.~K.}\ \bibnamefont {Malia}}, \bibinfo {author} {\bibfnamefont {T.}~\bibnamefont {Onodera}},\ and\ \bibinfo {author} {\bibfnamefont {P.~L.}\ \bibnamefont {McMahon}},\ }\href {http://arxiv.org/abs/2401.06119} {\bibinfo {title} {Highly multimode visible squeezed light with programmable spectral correlations through broadband up-conversion}} (\bibinfo {year} {2024}),\ \bibinfo {note} {arXiv:2401.06119 [physics, physics:quant-ph]}\BibitemShut {NoStop}%
\bibitem [{\citenamefont {Lustig}\ \emph {et~al.}(2025)\citenamefont {Lustig}, \citenamefont {Guidry}, \citenamefont {Lukin}, \citenamefont {Fan},\ and\ \citenamefont {Vuckovic}}]{lustig_quadrature-dependent_2025}%
  \BibitemOpen
  \bibfield  {author} {\bibinfo {author} {\bibfnamefont {E.}~\bibnamefont {Lustig}}, \bibinfo {author} {\bibfnamefont {M.~A.}\ \bibnamefont {Guidry}}, \bibinfo {author} {\bibfnamefont {D.~M.}\ \bibnamefont {Lukin}}, \bibinfo {author} {\bibfnamefont {S.}~\bibnamefont {Fan}},\ and\ \bibinfo {author} {\bibfnamefont {J.}~\bibnamefont {Vuckovic}},\ }\href {https://doi.org/10.48550/arXiv.2407.13049} {\bibinfo {title} {Quadrature-{Dependent} {Lattice} {Dynamics} of {Dissipative} {Microcombs}}} (\bibinfo {year} {2025}),\ \bibinfo {note} {arXiv:2407.13049 [physics]}\BibitemShut {NoStop}%
\bibitem [{\citenamefont {Zia~Uddin}\ \emph {et~al.}(2025)\citenamefont {Zia~Uddin}, \citenamefont {Rivera}, \citenamefont {Seyler}, \citenamefont {Sloan}, \citenamefont {Salamin}, \citenamefont {Roques-Carmes}, \citenamefont {Xu}, \citenamefont {Sander}, \citenamefont {Kaminer},\ and\ \citenamefont {Soljačić}}]{zia_uddin_noise-immune_2025}%
  \BibitemOpen
  \bibfield  {author} {\bibinfo {author} {\bibfnamefont {S.}~\bibnamefont {Zia~Uddin}}, \bibinfo {author} {\bibfnamefont {N.}~\bibnamefont {Rivera}}, \bibinfo {author} {\bibfnamefont {D.}~\bibnamefont {Seyler}}, \bibinfo {author} {\bibfnamefont {J.}~\bibnamefont {Sloan}}, \bibinfo {author} {\bibfnamefont {Y.}~\bibnamefont {Salamin}}, \bibinfo {author} {\bibfnamefont {C.}~\bibnamefont {Roques-Carmes}}, \bibinfo {author} {\bibfnamefont {S.}~\bibnamefont {Xu}}, \bibinfo {author} {\bibfnamefont {M.~Y.}\ \bibnamefont {Sander}}, \bibinfo {author} {\bibfnamefont {I.}~\bibnamefont {Kaminer}},\ and\ \bibinfo {author} {\bibfnamefont {M.}~\bibnamefont {Soljačić}},\ }\bibfield  {journal} {\bibinfo  {journal} {Nature Photonics}\ }\href {https://doi.org/10.1038/s41566-025-01677-2} {10.1038/s41566-025-01677-2} (\bibinfo {year} {2025})\BibitemShut {NoStop}%
\bibitem [{\citenamefont {Cai}\ \emph {et~al.}(2017)\citenamefont {Cai}, \citenamefont {Roslund}, \citenamefont {Ferrini}, \citenamefont {Arzani}, \citenamefont {Xu}, \citenamefont {Fabre},\ and\ \citenamefont {Treps}}]{cai_multimode_2017}%
  \BibitemOpen
  \bibfield  {author} {\bibinfo {author} {\bibfnamefont {Y.}~\bibnamefont {Cai}}, \bibinfo {author} {\bibfnamefont {J.}~\bibnamefont {Roslund}}, \bibinfo {author} {\bibfnamefont {G.}~\bibnamefont {Ferrini}}, \bibinfo {author} {\bibfnamefont {F.}~\bibnamefont {Arzani}}, \bibinfo {author} {\bibfnamefont {X.}~\bibnamefont {Xu}}, \bibinfo {author} {\bibfnamefont {C.}~\bibnamefont {Fabre}},\ and\ \bibinfo {author} {\bibfnamefont {N.}~\bibnamefont {Treps}},\ }\bibfield  {journal} {\bibinfo  {journal} {Nature Communications}\ }\textbf {\bibinfo {volume} {8}},\ \href {https://doi.org/10.1038/ncomms15645} {10.1038/ncomms15645} (\bibinfo {year} {2017}),\ \bibinfo {note} {publisher: Springer Science and Business Media LLC}\BibitemShut {NoStop}%
\bibitem [{\citenamefont {Asavanant}\ \emph {et~al.}(2019)\citenamefont {Asavanant}, \citenamefont {Shiozawa}, \citenamefont {Yokoyama}, \citenamefont {Charoensombutamon}, \citenamefont {Emura}, \citenamefont {Alexander}, \citenamefont {Takeda}, \citenamefont {Yoshikawa}, \citenamefont {Menicucci}, \citenamefont {Yonezawa},\ and\ \citenamefont {Furusawa}}]{asavanant_generation_2019}%
  \BibitemOpen
  \bibfield  {author} {\bibinfo {author} {\bibfnamefont {W.}~\bibnamefont {Asavanant}}, \bibinfo {author} {\bibfnamefont {Y.}~\bibnamefont {Shiozawa}}, \bibinfo {author} {\bibfnamefont {S.}~\bibnamefont {Yokoyama}}, \bibinfo {author} {\bibfnamefont {B.}~\bibnamefont {Charoensombutamon}}, \bibinfo {author} {\bibfnamefont {H.}~\bibnamefont {Emura}}, \bibinfo {author} {\bibfnamefont {R.~N.}\ \bibnamefont {Alexander}}, \bibinfo {author} {\bibfnamefont {S.}~\bibnamefont {Takeda}}, \bibinfo {author} {\bibfnamefont {J.-i.}\ \bibnamefont {Yoshikawa}}, \bibinfo {author} {\bibfnamefont {N.~C.}\ \bibnamefont {Menicucci}}, \bibinfo {author} {\bibfnamefont {H.}~\bibnamefont {Yonezawa}},\ and\ \bibinfo {author} {\bibfnamefont {A.}~\bibnamefont {Furusawa}},\ }\href {https://doi.org/10.1126/science.aay2645} {\bibfield  {journal} {\bibinfo  {journal} {Science}\ }\textbf {\bibinfo {volume} {366}},\ \bibinfo {pages} {373} (\bibinfo {year} {2019})}\BibitemShut {NoStop}%
\bibitem [{\citenamefont {Barral}\ \emph {et~al.}(2020)\citenamefont {Barral}, \citenamefont {Walschaers}, \citenamefont {Bencheikh}, \citenamefont {Parigi}, \citenamefont {Levenson}, \citenamefont {Treps},\ and\ \citenamefont {Belabas}}]{barral_versatile_2020}%
  \BibitemOpen
  \bibfield  {author} {\bibinfo {author} {\bibfnamefont {D.}~\bibnamefont {Barral}}, \bibinfo {author} {\bibfnamefont {M.}~\bibnamefont {Walschaers}}, \bibinfo {author} {\bibfnamefont {K.}~\bibnamefont {Bencheikh}}, \bibinfo {author} {\bibfnamefont {V.}~\bibnamefont {Parigi}}, \bibinfo {author} {\bibfnamefont {J.~A.}\ \bibnamefont {Levenson}}, \bibinfo {author} {\bibfnamefont {N.}~\bibnamefont {Treps}},\ and\ \bibinfo {author} {\bibfnamefont {N.}~\bibnamefont {Belabas}},\ }\href {https://doi.org/10.1103/PhysRevApplied.14.044025} {\bibfield  {journal} {\bibinfo  {journal} {Physical Review Applied}\ }\textbf {\bibinfo {volume} {14}},\ \bibinfo {pages} {044025} (\bibinfo {year} {2020})}\BibitemShut {NoStop}%
\bibitem [{\citenamefont {Nichols}\ \emph {et~al.}(2018)\citenamefont {Nichols}, \citenamefont {Liuzzo-Scorpo}, \citenamefont {Knott},\ and\ \citenamefont {Adesso}}]{nichols_multiparameter_2018}%
  \BibitemOpen
  \bibfield  {author} {\bibinfo {author} {\bibfnamefont {R.}~\bibnamefont {Nichols}}, \bibinfo {author} {\bibfnamefont {P.}~\bibnamefont {Liuzzo-Scorpo}}, \bibinfo {author} {\bibfnamefont {P.~A.}\ \bibnamefont {Knott}},\ and\ \bibinfo {author} {\bibfnamefont {G.}~\bibnamefont {Adesso}},\ }\href {https://doi.org/10.1103/PhysRevA.98.012114} {\bibfield  {journal} {\bibinfo  {journal} {Physical Review A}\ }\textbf {\bibinfo {volume} {98}},\ \bibinfo {pages} {012114} (\bibinfo {year} {2018})}\BibitemShut {NoStop}%
\bibitem [{\citenamefont {{The LIGO Scientific Collaboration}}(2011)}]{the_ligo_scientific_collaboration_gravitational_2011}%
  \BibitemOpen
  \bibfield  {author} {\bibinfo {author} {\bibnamefont {{The LIGO Scientific Collaboration}}},\ }\href {https://doi.org/10.1038/nphys2083} {\bibfield  {journal} {\bibinfo  {journal} {Nature Physics}\ }\textbf {\bibinfo {volume} {7}},\ \bibinfo {pages} {962} (\bibinfo {year} {2011})}\BibitemShut {NoStop}%
\bibitem [{\citenamefont {Brida}\ \emph {et~al.}(2010)\citenamefont {Brida}, \citenamefont {Genovese},\ and\ \citenamefont {Ruo~Berchera}}]{brida_experimental_2010}%
  \BibitemOpen
  \bibfield  {author} {\bibinfo {author} {\bibfnamefont {G.}~\bibnamefont {Brida}}, \bibinfo {author} {\bibfnamefont {M.}~\bibnamefont {Genovese}},\ and\ \bibinfo {author} {\bibfnamefont {I.}~\bibnamefont {Ruo~Berchera}},\ }\href {https://doi.org/10.1038/nphoton.2010.29} {\bibfield  {journal} {\bibinfo  {journal} {Nature Photonics}\ }\textbf {\bibinfo {volume} {4}},\ \bibinfo {pages} {227} (\bibinfo {year} {2010})}\BibitemShut {NoStop}%
\bibitem [{\citenamefont {Whittaker}\ \emph {et~al.}(2017)\citenamefont {Whittaker}, \citenamefont {Erven}, \citenamefont {Neville}, \citenamefont {Berry}, \citenamefont {O’Brien}, \citenamefont {Cable},\ and\ \citenamefont {Matthews}}]{whittaker_absorption_2017}%
  \BibitemOpen
  \bibfield  {author} {\bibinfo {author} {\bibfnamefont {R.}~\bibnamefont {Whittaker}}, \bibinfo {author} {\bibfnamefont {C.}~\bibnamefont {Erven}}, \bibinfo {author} {\bibfnamefont {A.}~\bibnamefont {Neville}}, \bibinfo {author} {\bibfnamefont {M.}~\bibnamefont {Berry}}, \bibinfo {author} {\bibfnamefont {J.~L.}\ \bibnamefont {O’Brien}}, \bibinfo {author} {\bibfnamefont {H.}~\bibnamefont {Cable}},\ and\ \bibinfo {author} {\bibfnamefont {J.~C.~F.}\ \bibnamefont {Matthews}},\ }\href {https://doi.org/10.1088/1367-2630/aa5512} {\bibfield  {journal} {\bibinfo  {journal} {New Journal of Physics}\ }\textbf {\bibinfo {volume} {19}},\ \bibinfo {pages} {023013} (\bibinfo {year} {2017})}\BibitemShut {NoStop}%
\bibitem [{\citenamefont {Ma}\ \emph {et~al.}(2025)\citenamefont {Ma}, \citenamefont {Wang}, \citenamefont {Laydevant}, \citenamefont {Wright},\ and\ \citenamefont {McMahon}}]{ma_quantum-limited_2025}%
  \BibitemOpen
  \bibfield  {author} {\bibinfo {author} {\bibfnamefont {S.-Y.}\ \bibnamefont {Ma}}, \bibinfo {author} {\bibfnamefont {T.}~\bibnamefont {Wang}}, \bibinfo {author} {\bibfnamefont {J.}~\bibnamefont {Laydevant}}, \bibinfo {author} {\bibfnamefont {L.~G.}\ \bibnamefont {Wright}},\ and\ \bibinfo {author} {\bibfnamefont {P.~L.}\ \bibnamefont {McMahon}},\ }\href {https://doi.org/10.1038/s41467-024-55220-y} {\bibfield  {journal} {\bibinfo  {journal} {Nature Communications}\ }\textbf {\bibinfo {volume} {16}},\ \bibinfo {pages} {359} (\bibinfo {year} {2025})}\BibitemShut {NoStop}%
\bibitem [{\citenamefont {Shapiro}(2009)}]{shapiro_quantum_2009}%
  \BibitemOpen
  \bibfield  {author} {\bibinfo {author} {\bibfnamefont {J.}~\bibnamefont {Shapiro}},\ }\href {https://doi.org/10.1109/JSTQE.2009.2024959} {\bibfield  {journal} {\bibinfo  {journal} {IEEE Journal of Selected Topics in Quantum Electronics}\ }\textbf {\bibinfo {volume} {15}},\ \bibinfo {pages} {1547} (\bibinfo {year} {2009})}\BibitemShut {NoStop}%
\bibitem [{\citenamefont {Weedbrook}\ \emph {et~al.}(2012)\citenamefont {Weedbrook}, \citenamefont {Pirandola}, \citenamefont {García-Patrón}, \citenamefont {Cerf}, \citenamefont {Ralph}, \citenamefont {Shapiro},\ and\ \citenamefont {Lloyd}}]{weedbrook_gaussian_2012}%
  \BibitemOpen
  \bibfield  {author} {\bibinfo {author} {\bibfnamefont {C.}~\bibnamefont {Weedbrook}}, \bibinfo {author} {\bibfnamefont {S.}~\bibnamefont {Pirandola}}, \bibinfo {author} {\bibfnamefont {R.}~\bibnamefont {García-Patrón}}, \bibinfo {author} {\bibfnamefont {N.~J.}\ \bibnamefont {Cerf}}, \bibinfo {author} {\bibfnamefont {T.~C.}\ \bibnamefont {Ralph}}, \bibinfo {author} {\bibfnamefont {J.~H.}\ \bibnamefont {Shapiro}},\ and\ \bibinfo {author} {\bibfnamefont {S.}~\bibnamefont {Lloyd}},\ }\href {https://doi.org/10.1103/RevModPhys.84.621} {\bibfield  {journal} {\bibinfo  {journal} {Reviews of Modern Physics}\ }\textbf {\bibinfo {volume} {84}},\ \bibinfo {pages} {621} (\bibinfo {year} {2012})}\BibitemShut {NoStop}%
\bibitem [{\citenamefont {Loudon}(2000)}]{loudon_quantum_2000}%
  \BibitemOpen
  \bibfield  {author} {\bibinfo {author} {\bibfnamefont {R.}~\bibnamefont {Loudon}},\ }\href {https://doi.org/10.1093/oso/9780198501770.001.0001} {{\selectlanguage {english}\emph {\bibinfo {title} {The {Quantum} {Theory} of {Light}}}}},\ \bibinfo {edition} {third edition}\ ed.\ (\bibinfo  {publisher} {Oxford University Press},\ \bibinfo {year} {2000})\BibitemShut {NoStop}%
\bibitem [{\citenamefont {Bachor}\ and\ \citenamefont {Ralph}(2019)}]{bachor_guide_2019}%
  \BibitemOpen
  \bibfield  {author} {\bibinfo {author} {\bibfnamefont {H.-A.}\ \bibnamefont {Bachor}}\ and\ \bibinfo {author} {\bibfnamefont {T.~C.}\ \bibnamefont {Ralph}},\ }\href {https://doi.org/10.1002/9783527695805} {{\selectlanguage {english}\emph {\bibinfo {title} {A guide to experiments in quantum optics}}}},\ \bibinfo {edition} {3rd}\ ed.\ (\bibinfo  {publisher} {Wiley-VCH},\ \bibinfo {year} {2019})\BibitemShut {NoStop}%
\bibitem [{\citenamefont {Corwin}\ \emph {et~al.}(2003)\citenamefont {Corwin}, \citenamefont {Newbury}, \citenamefont {Dudley}, \citenamefont {Coen}, \citenamefont {Diddams}, \citenamefont {Weber},\ and\ \citenamefont {Windeler}}]{corwin_fundamental_2003}%
  \BibitemOpen
  \bibfield  {author} {\bibinfo {author} {\bibfnamefont {K.~L.}\ \bibnamefont {Corwin}}, \bibinfo {author} {\bibfnamefont {N.~R.}\ \bibnamefont {Newbury}}, \bibinfo {author} {\bibfnamefont {J.~M.}\ \bibnamefont {Dudley}}, \bibinfo {author} {\bibfnamefont {S.}~\bibnamefont {Coen}}, \bibinfo {author} {\bibfnamefont {S.~A.}\ \bibnamefont {Diddams}}, \bibinfo {author} {\bibfnamefont {K.}~\bibnamefont {Weber}},\ and\ \bibinfo {author} {\bibfnamefont {R.~S.}\ \bibnamefont {Windeler}},\ }\href {https://doi.org/10.1103/PhysRevLett.90.113904} {\bibfield  {journal} {\bibinfo  {journal} {Physical Review Letters}\ }\textbf {\bibinfo {volume} {90}},\ \bibinfo {pages} {113904} (\bibinfo {year} {2003})}\BibitemShut {NoStop}%
\bibitem [{Note1()}]{Note1}%
  \BibitemOpen
  \bibinfo {note} {We consider the power spectral density at a given frequency rather than the frequency-integrated power spectral density (which is proportional to the photon number variance in a pulse), as this allows us to make direct comparisons with the quantum shot noise level (as is standard). In contrast, the integrated power spectral density will typically be far from the shot noise level due to low-frequency technical noise sources --- however this noise can also be reduced by the same methods described here due to the quasi-instantaneous nature of Kerr nonlinearity.}\BibitemShut {Stop}%
\bibitem [{\citenamefont {Cavalcanti}\ \emph {et~al.}(1995)\citenamefont {Cavalcanti}, \citenamefont {Agrawal},\ and\ \citenamefont {Yu}}]{cavalcanti_noise_1995}%
  \BibitemOpen
  \bibfield  {author} {\bibinfo {author} {\bibfnamefont {S.~B.}\ \bibnamefont {Cavalcanti}}, \bibinfo {author} {\bibfnamefont {G.~P.}\ \bibnamefont {Agrawal}},\ and\ \bibinfo {author} {\bibfnamefont {M.}~\bibnamefont {Yu}},\ }\href {https://doi.org/10.1103/PhysRevA.51.4086} {\bibfield  {journal} {\bibinfo  {journal} {Physical Review A}\ }\textbf {\bibinfo {volume} {51}},\ \bibinfo {pages} {4086} (\bibinfo {year} {1995})}\BibitemShut {NoStop}%
\bibitem [{\citenamefont {Caves}(1982)}]{caves_quantum_1982}%
  \BibitemOpen
  \bibfield  {author} {\bibinfo {author} {\bibfnamefont {C.~M.}\ \bibnamefont {Caves}},\ }\href {https://doi.org/10.1103/PhysRevD.26.1817} {\bibfield  {journal} {\bibinfo  {journal} {Physical Review D}\ }\textbf {\bibinfo {volume} {26}},\ \bibinfo {pages} {1817} (\bibinfo {year} {1982})}\BibitemShut {NoStop}%
\bibitem [{\citenamefont {Walmsley}\ and\ \citenamefont {Raymer}(1983)}]{walmsley_observation_1983}%
  \BibitemOpen
  \bibfield  {author} {\bibinfo {author} {\bibfnamefont {I.~A.}\ \bibnamefont {Walmsley}}\ and\ \bibinfo {author} {\bibfnamefont {M.~G.}\ \bibnamefont {Raymer}},\ }\href {https://doi.org/10.1103/PhysRevLett.50.962} {\bibfield  {journal} {\bibinfo  {journal} {Physical Review Letters}\ }\textbf {\bibinfo {volume} {50}},\ \bibinfo {pages} {962} (\bibinfo {year} {1983})}\BibitemShut {NoStop}%
\bibitem [{\citenamefont {Dudley}\ \emph {et~al.}(2006)\citenamefont {Dudley}, \citenamefont {Genty},\ and\ \citenamefont {Coen}}]{dudley_supercontinuum_2006}%
  \BibitemOpen
  \bibfield  {author} {\bibinfo {author} {\bibfnamefont {J.~M.}\ \bibnamefont {Dudley}}, \bibinfo {author} {\bibfnamefont {G.}~\bibnamefont {Genty}},\ and\ \bibinfo {author} {\bibfnamefont {S.}~\bibnamefont {Coen}},\ }\href {https://doi.org/10.1103/RevModPhys.78.1135} {\bibfield  {journal} {\bibinfo  {journal} {Reviews of Modern Physics}\ }\textbf {\bibinfo {volume} {78}},\ \bibinfo {pages} {1135} (\bibinfo {year} {2006})}\BibitemShut {NoStop}%
\bibitem [{\citenamefont {Gross}\ and\ \citenamefont {Haroche}(1982)}]{gross_superradiance_1982}%
  \BibitemOpen
  \bibfield  {author} {\bibinfo {author} {\bibfnamefont {M.}~\bibnamefont {Gross}}\ and\ \bibinfo {author} {\bibfnamefont {S.}~\bibnamefont {Haroche}},\ }\href {https://doi.org/10.1016/0370-1573(82)90102-8} {\bibfield  {journal} {\bibinfo  {journal} {Physics Reports}\ }\textbf {\bibinfo {volume} {93}},\ \bibinfo {pages} {301} (\bibinfo {year} {1982})}\BibitemShut {NoStop}%
\bibitem [{\citenamefont {Roques-Carmes}\ \emph {et~al.}(2023)\citenamefont {Roques-Carmes}, \citenamefont {Salamin}, \citenamefont {Sloan}, \citenamefont {Choi}, \citenamefont {Velez}, \citenamefont {Koskas}, \citenamefont {Rivera}, \citenamefont {Kooi}, \citenamefont {Joannopoulos},\ and\ \citenamefont {Soljačić}}]{roques-carmes_biasing_2023}%
  \BibitemOpen
  \bibfield  {author} {\bibinfo {author} {\bibfnamefont {C.}~\bibnamefont {Roques-Carmes}}, \bibinfo {author} {\bibfnamefont {Y.}~\bibnamefont {Salamin}}, \bibinfo {author} {\bibfnamefont {J.}~\bibnamefont {Sloan}}, \bibinfo {author} {\bibfnamefont {S.}~\bibnamefont {Choi}}, \bibinfo {author} {\bibfnamefont {G.}~\bibnamefont {Velez}}, \bibinfo {author} {\bibfnamefont {E.}~\bibnamefont {Koskas}}, \bibinfo {author} {\bibfnamefont {N.}~\bibnamefont {Rivera}}, \bibinfo {author} {\bibfnamefont {S.~E.}\ \bibnamefont {Kooi}}, \bibinfo {author} {\bibfnamefont {J.~D.}\ \bibnamefont {Joannopoulos}},\ and\ \bibinfo {author} {\bibfnamefont {M.}~\bibnamefont {Soljačić}},\ }\href {https://doi.org/10.1126/science.adh4920} {\bibfield  {journal} {\bibinfo  {journal} {Science}\ }\textbf {\bibinfo {volume} {381}},\ \bibinfo {pages} {205} (\bibinfo {year} {2023})}\BibitemShut {NoStop}%
\bibitem [{\citenamefont {Pellegrini}\ \emph {et~al.}(2016)\citenamefont {Pellegrini}, \citenamefont {Marinelli},\ and\ \citenamefont {Reiche}}]{pellegrini_physics_2016}%
  \BibitemOpen
  \bibfield  {author} {\bibinfo {author} {\bibfnamefont {C.}~\bibnamefont {Pellegrini}}, \bibinfo {author} {\bibfnamefont {A.}~\bibnamefont {Marinelli}},\ and\ \bibinfo {author} {\bibfnamefont {S.}~\bibnamefont {Reiche}},\ }\href {https://doi.org/10.1103/RevModPhys.88.015006} {\bibfield  {journal} {\bibinfo  {journal} {Reviews of Modern Physics}\ }\textbf {\bibinfo {volume} {88}},\ \bibinfo {pages} {015006} (\bibinfo {year} {2016})}\BibitemShut {NoStop}%
\bibitem [{\citenamefont {Cao}\ \emph {et~al.}(2022)\citenamefont {Cao}, \citenamefont {Mosk},\ and\ \citenamefont {Rotter}}]{cao_shaping_2022}%
  \BibitemOpen
  \bibfield  {author} {\bibinfo {author} {\bibfnamefont {H.}~\bibnamefont {Cao}}, \bibinfo {author} {\bibfnamefont {A.~P.}\ \bibnamefont {Mosk}},\ and\ \bibinfo {author} {\bibfnamefont {S.}~\bibnamefont {Rotter}},\ }\href {https://doi.org/10.1038/s41567-022-01677-x} {\bibfield  {journal} {\bibinfo  {journal} {Nature Physics}\ }\textbf {\bibinfo {volume} {18}},\ \bibinfo {pages} {994} (\bibinfo {year} {2022})}\BibitemShut {NoStop}%
\bibitem [{\citenamefont {Tzang}\ \emph {et~al.}(2018)\citenamefont {Tzang}, \citenamefont {Caravaca-Aguirre}, \citenamefont {Wagner},\ and\ \citenamefont {Piestun}}]{tzang_adaptive_2018}%
  \BibitemOpen
  \bibfield  {author} {\bibinfo {author} {\bibfnamefont {O.}~\bibnamefont {Tzang}}, \bibinfo {author} {\bibfnamefont {A.~M.}\ \bibnamefont {Caravaca-Aguirre}}, \bibinfo {author} {\bibfnamefont {K.}~\bibnamefont {Wagner}},\ and\ \bibinfo {author} {\bibfnamefont {R.}~\bibnamefont {Piestun}},\ }\href {https://doi.org/10.1038/s41566-018-0167-7} {\bibfield  {journal} {\bibinfo  {journal} {Nature Photonics}\ }\textbf {\bibinfo {volume} {12}},\ \bibinfo {pages} {368} (\bibinfo {year} {2018})}\BibitemShut {NoStop}%
\bibitem [{\citenamefont {Florentin}\ \emph {et~al.}(2016)\citenamefont {Florentin}, \citenamefont {Kermene}, \citenamefont {Benoist}, \citenamefont {Desfarges-Berthelemot}, \citenamefont {Pagnoux}, \citenamefont {Barthélémy},\ and\ \citenamefont {Huignard}}]{florentin_shaping_2016}%
  \BibitemOpen
  \bibfield  {author} {\bibinfo {author} {\bibfnamefont {R.}~\bibnamefont {Florentin}}, \bibinfo {author} {\bibfnamefont {V.}~\bibnamefont {Kermene}}, \bibinfo {author} {\bibfnamefont {J.}~\bibnamefont {Benoist}}, \bibinfo {author} {\bibfnamefont {A.}~\bibnamefont {Desfarges-Berthelemot}}, \bibinfo {author} {\bibfnamefont {D.}~\bibnamefont {Pagnoux}}, \bibinfo {author} {\bibfnamefont {A.}~\bibnamefont {Barthélémy}},\ and\ \bibinfo {author} {\bibfnamefont {J.-P.}\ \bibnamefont {Huignard}},\ }\href {https://doi.org/10.1038/lsa.2016.208} {\bibfield  {journal} {\bibinfo  {journal} {Light: Science \& Applications}\ }\textbf {\bibinfo {volume} {6}},\ \bibinfo {pages} {e16208} (\bibinfo {year} {2016})}\BibitemShut {NoStop}%
\bibitem [{\citenamefont {Wisal}\ \emph {et~al.}(2024)\citenamefont {Wisal}, \citenamefont {Chen}, \citenamefont {Kuang}, \citenamefont {Miller}, \citenamefont {Cao},\ and\ \citenamefont {Stone}}]{wisal_optimal_2024}%
  \BibitemOpen
  \bibfield  {author} {\bibinfo {author} {\bibfnamefont {K.}~\bibnamefont {Wisal}}, \bibinfo {author} {\bibfnamefont {C.-W.}\ \bibnamefont {Chen}}, \bibinfo {author} {\bibfnamefont {Z.}~\bibnamefont {Kuang}}, \bibinfo {author} {\bibfnamefont {O.~D.}\ \bibnamefont {Miller}}, \bibinfo {author} {\bibfnamefont {H.}~\bibnamefont {Cao}},\ and\ \bibinfo {author} {\bibfnamefont {A.~D.}\ \bibnamefont {Stone}},\ }\href {http://arxiv.org/abs/2407.05201} {\bibinfo {title} {Optimal input excitations for suppressing nonlinear instabilities in multimode fibers}} (\bibinfo {year} {2024}),\ \bibinfo {note} {arXiv:2407.05201 [physics]}\BibitemShut {NoStop}%
\bibitem [{\citenamefont {Boussafa}\ \emph {et~al.}(2025)\citenamefont {Boussafa}, \citenamefont {Sader}, \citenamefont {Hoang}, \citenamefont {Chaves}, \citenamefont {Bougaud}, \citenamefont {Fabert}, \citenamefont {Tonello}, \citenamefont {Dudley}, \citenamefont {Kues},\ and\ \citenamefont {Wetzel}}]{boussafa_deep_2025}%
  \BibitemOpen
  \bibfield  {author} {\bibinfo {author} {\bibfnamefont {Y.}~\bibnamefont {Boussafa}}, \bibinfo {author} {\bibfnamefont {L.}~\bibnamefont {Sader}}, \bibinfo {author} {\bibfnamefont {V.~T.}\ \bibnamefont {Hoang}}, \bibinfo {author} {\bibfnamefont {B.~P.}\ \bibnamefont {Chaves}}, \bibinfo {author} {\bibfnamefont {A.}~\bibnamefont {Bougaud}}, \bibinfo {author} {\bibfnamefont {M.}~\bibnamefont {Fabert}}, \bibinfo {author} {\bibfnamefont {A.}~\bibnamefont {Tonello}}, \bibinfo {author} {\bibfnamefont {J.~M.}\ \bibnamefont {Dudley}}, \bibinfo {author} {\bibfnamefont {M.}~\bibnamefont {Kues}},\ and\ \bibinfo {author} {\bibfnamefont {B.}~\bibnamefont {Wetzel}},\ }\href {https://doi.org/10.1038/s41467-025-62713-x} {\bibfield  {journal} {\bibinfo  {journal} {Nature Communications}\ }\textbf {\bibinfo {volume} {16}},\ \bibinfo {pages} {7800} (\bibinfo {year} {2025})}\BibitemShut {NoStop}%
\bibitem [{\citenamefont {Chen}\ \emph {et~al.}(2023{\natexlab{b}})\citenamefont {Chen}, \citenamefont {Nguyen}, \citenamefont {Wisal}, \citenamefont {Wei}, \citenamefont {Warren-Smith}, \citenamefont {Henderson-Sapir}, \citenamefont {Schartner}, \citenamefont {Ahmadi}, \citenamefont {Ebendorff-Heidepriem}, \citenamefont {Stone}, \citenamefont {Ottaway},\ and\ \citenamefont {Cao}}]{chen_mitigating_2023}%
  \BibitemOpen
  \bibfield  {author} {\bibinfo {author} {\bibfnamefont {C.-W.}\ \bibnamefont {Chen}}, \bibinfo {author} {\bibfnamefont {L.~V.}\ \bibnamefont {Nguyen}}, \bibinfo {author} {\bibfnamefont {K.}~\bibnamefont {Wisal}}, \bibinfo {author} {\bibfnamefont {S.}~\bibnamefont {Wei}}, \bibinfo {author} {\bibfnamefont {S.~C.}\ \bibnamefont {Warren-Smith}}, \bibinfo {author} {\bibfnamefont {O.}~\bibnamefont {Henderson-Sapir}}, \bibinfo {author} {\bibfnamefont {E.~P.}\ \bibnamefont {Schartner}}, \bibinfo {author} {\bibfnamefont {P.}~\bibnamefont {Ahmadi}}, \bibinfo {author} {\bibfnamefont {H.}~\bibnamefont {Ebendorff-Heidepriem}}, \bibinfo {author} {\bibfnamefont {A.~D.}\ \bibnamefont {Stone}}, \bibinfo {author} {\bibfnamefont {D.~J.}\ \bibnamefont {Ottaway}},\ and\ \bibinfo {author} {\bibfnamefont {H.}~\bibnamefont {Cao}},\ }\href {https://doi.org/10.1038/s41467-023-42806-1} {\bibfield  {journal} {\bibinfo  {journal} {Nature Communications}\ }\textbf {\bibinfo {volume} {14}},\ \bibinfo {pages} {7343} (\bibinfo {year}
  {2023}{\natexlab{b}})}\BibitemShut {NoStop}%
\bibitem [{\citenamefont {Wisal}\ \emph {et~al.}(2023)\citenamefont {Wisal}, \citenamefont {Warren-Smith}, \citenamefont {Chen}, \citenamefont {Cao},\ and\ \citenamefont {Stone}}]{wisal_theory_2023}%
  \BibitemOpen
  \bibfield  {author} {\bibinfo {author} {\bibfnamefont {K.}~\bibnamefont {Wisal}}, \bibinfo {author} {\bibfnamefont {S.~C.}\ \bibnamefont {Warren-Smith}}, \bibinfo {author} {\bibfnamefont {C.-W.}\ \bibnamefont {Chen}}, \bibinfo {author} {\bibfnamefont {H.}~\bibnamefont {Cao}},\ and\ \bibinfo {author} {\bibfnamefont {A.~D.}\ \bibnamefont {Stone}},\ }\href {http://arxiv.org/abs/2304.09342} {\bibinfo {title} {Theory of {Stimulated} {Brillouin} {Scattering} in {Fibers} for {Highly} {Multimode} {Excitations}}} (\bibinfo {year} {2023}),\ \bibinfo {note} {arXiv:2304.09342 [physics]}\BibitemShut {NoStop}%
\bibitem [{\citenamefont {Agrawal}(2019)}]{agrawal_invite_2019}%
  \BibitemOpen
  \bibfield  {author} {\bibinfo {author} {\bibfnamefont {G.~P.}\ \bibnamefont {Agrawal}},\ }\href {https://doi.org/10.1016/j.yofte.2019.04.012} {\bibfield  {journal} {\bibinfo  {journal} {Optical Fiber Technology}\ }\textbf {\bibinfo {volume} {50}},\ \bibinfo {pages} {309} (\bibinfo {year} {2019})}\BibitemShut {NoStop}%
\bibitem [{\citenamefont {Wright}\ \emph {et~al.}(2015)\citenamefont {Wright}, \citenamefont {Christodoulides},\ and\ \citenamefont {Wise}}]{wright_controllable_2015}%
  \BibitemOpen
  \bibfield  {author} {\bibinfo {author} {\bibfnamefont {L.~G.}\ \bibnamefont {Wright}}, \bibinfo {author} {\bibfnamefont {D.~N.}\ \bibnamefont {Christodoulides}},\ and\ \bibinfo {author} {\bibfnamefont {F.~W.}\ \bibnamefont {Wise}},\ }\href {https://doi.org/10.1038/nphoton.2015.61} {\bibfield  {journal} {\bibinfo  {journal} {Nature Photonics}\ }\textbf {\bibinfo {volume} {9}},\ \bibinfo {pages} {306} (\bibinfo {year} {2015})}\BibitemShut {NoStop}%
\bibitem [{\citenamefont {Richardson}\ \emph {et~al.}(2010)\citenamefont {Richardson}, \citenamefont {Nilsson},\ and\ \citenamefont {Clarkson}}]{richardson_high_2010}%
  \BibitemOpen
  \bibfield  {author} {\bibinfo {author} {\bibfnamefont {D.~J.}\ \bibnamefont {Richardson}}, \bibinfo {author} {\bibfnamefont {J.}~\bibnamefont {Nilsson}},\ and\ \bibinfo {author} {\bibfnamefont {W.~A.}\ \bibnamefont {Clarkson}},\ }\href {https://doi.org/10.1364/JOSAB.27.000B63} {\bibfield  {journal} {\bibinfo  {journal} {Journal of the Optical Society of America B}\ }\textbf {\bibinfo {volume} {27}},\ \bibinfo {pages} {B63} (\bibinfo {year} {2010})}\BibitemShut {NoStop}%
\bibitem [{\citenamefont {Gonzalez-Ballestero}\ \emph {et~al.}(2023)\citenamefont {Gonzalez-Ballestero}, \citenamefont {Zielińska}, \citenamefont {Rossi}, \citenamefont {Militaru}, \citenamefont {Frimmer}, \citenamefont {Novotny}, \citenamefont {Maurer},\ and\ \citenamefont {Romero-Isart}}]{gonzalez-ballestero_suppressing_2023}%
  \BibitemOpen
  \bibfield  {author} {\bibinfo {author} {\bibfnamefont {C.}~\bibnamefont {Gonzalez-Ballestero}}, \bibinfo {author} {\bibfnamefont {J.}~\bibnamefont {Zielińska}}, \bibinfo {author} {\bibfnamefont {M.}~\bibnamefont {Rossi}}, \bibinfo {author} {\bibfnamefont {A.}~\bibnamefont {Militaru}}, \bibinfo {author} {\bibfnamefont {M.}~\bibnamefont {Frimmer}}, \bibinfo {author} {\bibfnamefont {L.}~\bibnamefont {Novotny}}, \bibinfo {author} {\bibfnamefont {P.}~\bibnamefont {Maurer}},\ and\ \bibinfo {author} {\bibfnamefont {O.}~\bibnamefont {Romero-Isart}},\ }\href {https://doi.org/10.1103/PRXQuantum.4.030331} {\bibfield  {journal} {\bibinfo  {journal} {PRX Quantum}\ }\textbf {\bibinfo {volume} {4}},\ \bibinfo {pages} {030331} (\bibinfo {year} {2023})}\BibitemShut {NoStop}%
\end{thebibliography}

%

\end{document}